\documentclass{aaEC}

\usepackage{graphicx}
\usepackage{natbib}
\usepackage{scalerel}
\usepackage{comment}

\usepackage[table]{xcolor}

\usepackage{txfonts}
\usepackage[pdfencoding=auto,psdextra]{hyperref}
\hypersetup{
    colorlinks=true,
    linkcolor=blue,
    filecolor=magenta,      
    urlcolor=blue,
    citecolor=blue
}
\makeatletter
\renewcommand*\aa@pageof{, page \thepage{} of \pageref*{LastPage}}
\makeatother

\usepackage[utf8]{inputenc}

\usepackage[switch, modulo]{lineno}

\usepackage{euclid}

\usepackage[wide, rightcaption]{sidecap}   
\sidecaptionvpos{figure}{t}

\usepackage{tikz}
\usetikzlibrary{calc}
\newcommand*{\firstdir}{40}%
\newcommand*{\seconddir}{80}%
\newcommand*{\thirddir}{135}%
\newcommand*{\length}{4.5}%
\tikzset{>=latex}

\newcommand{\dd}{\mathrm{d}}
\newcommand{\eexp}{\mathrm{e}}
\newcommand{\iunit}{\mathrm{i}}
\newcommand{\Nap}{\mathcal{N}}
\newcommand{\Dlog}{\Delta_\mathrm{ln}}
\newcommand{\Dphi}{\Delta_\phi/2\pi}
\newcommand{\Aap}{A_\mathrm{ap}}
\newcommand{\epst}[1]{\epsilon_{\mathrm{t}#1}}
\newcommand{\thap}[2]{\theta_{\mathrm{#1}#2}}
\newcommand{\Naps}[1]{\mathcal{N}_{\mathrm{s},#1}}
\newcommand{\Maps}[1]{M_{\mathrm{s},#1}}

\newcommand{\Nlens}{\Bar{N}_\mathrm{L}}

\newcommand{\nnngCF}{G_\mathrm{ggg\gamma}}
\newcommand{\ggnnCF}{G_\mathrm{\gamma\gamma gg}}
\newcommand{\gggnCF}{G_\mathrm{\gamma\gamma\gamma g}}

\begin{document}
%
%

\title{Fourth-order galaxy-galaxy-lensing: Theoretical framework and direct estimation}

   
\newcommand{\orcid}[1]{} 
\author{Jonathan Oel\inst{1,2} \and Lucas Porth\inst{1} \and Peter Schneider\inst{1} \and
  Elena Silvestre-Rosello\inst{3,1}}

%
%
%
%

\institute{Argelander-Institut für Astronomie, University of Bonn, Auf dem Hügel 71, 53121 Bonn, Germany \and Astronomy Centre, Department of Physics and Astronomy, University of Sussex, Brighton, BN1 9RH, UK \and
Universität Innsbruck, Institut für Astro- und Teilchenphysik, Technikerstr. 25/8, 6020 Innsbruck, Austria \\
\email{J.J.Oel@sussex.ac.uk}
}

%
%
\abstract{
\textsl{Context.~}
Traditional galaxy-galaxy lensing is a well-established method of probing the statistical properties of the Universe's matter and galaxy distribution. 
However, this measure does not carry all the statistical information, provided the matter and galaxy distribution contain non-Gaussian features.
In order to study these non-Gaussianities, it is necessary to consider higher-order statistical measures. 

\textsl{Aims.~}
The aim of this work is to extend the analytical basis describing the statistical correlations between galaxies and shear to the fourth order, with special emphasis on the associated aperture statistics. 
In order to include fourth-order statistics in future analysis of the relation between mass and galaxies, we further investigate whether we can expect to detect these statistics from observations of stage IV surveys.

\textsl{Methods.~}
We define the four-point correlation function (4PCF) between the shear and the positions of triplets of foreground galaxies and derive its relation to the respective trispectrum. 
We convert the 4PCF to aperture statistics and derive the analytical form of the respective filter function, which we then implement in a numerical integration pipeline.
Furthermore, we develop a direct estimator that allows us to measure galaxy-mass aperture moments of arbitrary order on pixelized data using a Fast-Fourier-Transform (FFT) algorithm. 

\textsl{Results.~}
We show that the aperture measure $\langle\Nap^3\Map\rangle$ can be calculated with sub-percent accuracy on relevant aperture scales, $\theta$, by means of numerical integration.
Furthermore, we apply the FFT-based direct estimator to a mock catalog with a realistic stage IV survey setup on a sky area of $2000~\deg^2$, and detect the connected part of $\langle\Nap^3\Map\rangle(\theta)$ with a signal-to-noise ratio of roughly nine on small aperture scales.
}

%
%
    \keywords{Gravitational lensing: weak -- large-scale structure of Universe -- Methods: analytical -- Methods: numerical}
%
%
   \titlerunning{Fourth-order galaxy-galaxy-lensing}
   \authorrunning{Jonathan Oel et al.}
   
   \maketitle
%
%
%
%
   
\section{\label{sc:Intro}Introduction}
The constraints on cosmological parameters and the validity of any cosmological model rely on the detailed investigation of the properties of the cosmic microwave background and the large-scale distribution of matter and galaxies in the Universe. 
One of the well-established means to probe the statistical properties of the large-scale structure (LSS) is provided by cosmic shear  \citep[e.g.][]{Blandford_1991,Jain_1997,Schneider_1998b}, which describes the weak gravitational lensing effect caused by the inhomogeneities in the distribution of matter. 
Measurements of correlations in the cosmic shear field, as probed with the observed shapes of distant background galaxies, have mostly concentrated on second-order statistics \citep[e.g.][]{Bacon_2000, vanwaerbeke_2000}. 
The further technological advancement of cosmic shear surveys goes hand in hand with the significant detection of higher-order statistics \citep[e.g.][]{Bernardeau_2002,Porth_2024}, which are crucial in order to quantify the non-Gaussian information contained in the matter and galaxy distributions. 

Apart from shear correlations, one can also study the so-called galaxy-galaxy lensing (GGL) statistics which are described by the two-point correlation function between the positions of foreground ``lens'' galaxies with the shear field as measured from the shapes of background ``source'' galaxies. 
First detected by \cite{Brainerd_1996}, measurements of GGL have been performed for numerous surveys \citep[e.g.][]{Mandelbaum_2006,Heymans_2021} and can be used as a complementary probe to cosmic shear analyses to retrieve cosmological information \citep{Abbott_2022}.
Since galaxies are biased tracers of the underlying matter distribution, GGL quantifies the relation between the distributions of galaxies and dark matter, which can be modeled by a scale-dependent bias parameter \citep{Schneider_1998}. 
\cite{Schneider_2005} extended the notion of GGL to third-order galaxy-galaxy-galaxy lensing (G3L) statistics, first detected by \cite{Simon_2008} in the correlation between two lens galaxies and one source ellipticity in the Red-Sequence Cluster Survey \citep{Gladders_2005}. 
Since then, G3L has been subject to further studies in which it was shown that one can probe galaxy-mass correlations in higher order \citep{Simon_2013}, that by using a halo model of the galaxy-mass bispectrum, G3L can constrain the halo occupation distributions of galaxies \citep{Linke_2022}, and that G3L can put to test semi-analytic models of galaxy evolution \citep{Saghiha_2017,Linke_2020}. 

Ongoing stage IV surveys like \textit{Euclid} \citep{EuclidSkyOverview} and LSST \citep{Ivezi__2019} will observe cosmic shear at a new level in terms of depth and sky coverage. 
Therefore, in the further advancement of weak gravitational lensing as a tool to probe the properties of the LSS, it is important to assess whether we can expect statistical measures beyond the third order to be significantly detectable. 
In particular, by taking into account fourth-order correlations between galaxies and shear (G4L), we introduce a measure that is able to quantify the relation between mass and galaxies beyond the scope of GGL and G3L.

The weak lensing signal measured by GGL is sensitive to the mass environment of individual foreground galaxies, which is dominated by the properties of the galaxy dark matter halo on relatively small scales, and the general mass distribution of the group or cluster a galaxy is embedded in towards larger scales.
This allows us to study the, generally scale-dependent, bias parameter, $b$, and galaxy-mass correlation coefficient, $r$, which relate the matter power spectrum to the cross-power spectrum between galaxies and matter. 
For second-order GGL, aperture statistics present a suitable means for probing these parameters \citep[e.g.][]{Schneider_1998,Hoekstra_2001,Hoekstra_2002,Simon_2018,Simon_2021}. 
Therefore, throughout this work, we particularly focus on fourth-order aperture statistics that quantify the G4L signal. 
By considering higher-order correlations (i.e. G3L and G4L), one can further study the weak lensing signal induced by the mass environment around pairs and triplets of foreground galaxies, and therefore study the bias parameter and galaxy-mass correlation coefficient to higher order.

Higher-order galaxy-galaxy lensing signals are also sensitive to the more subtle properties of dark matter halos, like their overall shape -- for instance in a lens-shear-shear correlation -- and a possible misalignment between the profiles of the mass distribution and the brightness due to a preferred orientation of lens pairs, triplets, and so on \citep{Simon_2012}. 
However, since these applications require more thorough lens galaxy selections, here we only focus on forecasting the detection of G4L in the context of measuring $b$ and $r$. 


This work focuses on the development of the G4L statistics as the extension of the second- and third-order counterparts. 
First, in Sect. \ref{sc:theory} we give a brief overview of the concepts of weak lensing that are required to study G4L. 
In Sect. \ref{sc:fourth-order_theory}, we then introduce the definition of the four-point correlation function (4PCF) in G4L as well as the corresponding trispectrum, and the relations between these two quantities. 
Here we also consider the conversion of the 4PCF into aperture statistics and derive the corresponding formulas. 
In Sect. \ref{sc:measureG4L}, we develop a numerical integration pipeline in order to measure the fourth-order aperture statistics $\langle\Nap^3\Map\rangle$ from the measurement of the corresponding 4PCF, assuming Gaussian shear and galaxy fields.
In Sect. \ref{sc:estimation_apstats}, we present a direct estimator for general mixed aperture moments $\langle\Nap^n\Map^m\rangle$, based on the pixelization of the corresponding shear and galaxy fields. 
In order to assess the detection significance of aperture statistics, we apply this direct estimator to simulated data with a realistic stage IV survey cosmology setup and present the results in Sect. \ref{sc:detectability}.




\section{\label{sc:theory} Aperture measures for the relation between cosmic shear and the distribution of galaxies }
In this section, we briefly introduce the necessary concepts that are required to study fourth-order statistics in weak gravitational lensing. For a more detailed overview, see, e.g. \cite{Bartelmann_Schneider_2001} and \cite{Kilbinger_2015}.

\subsection{\label{ssc:weak_lensing} Weak lensing basics}
In modern cosmology, weak gravitational lensing is an important tool for probing the matter content of the Universe. 
Although causing only a rather subtle distortion of the observed image of a single distant galaxy, this effect becomes noticeable as a systematic image distortion in a larger ensemble of observed galaxy images. 
The observed galaxy shapes trace the underlying complex shear, $\gamma$, that emerges from the same scalar lensing potential as the convergence, $\kappa$.  
The distribution of matter as a function of the comoving distance, $\chi$, and the projected comoving coordinates, $\pmb{x}=f_K(\chi)~\pmb{\theta}$, with the comoving angular diameter distance, $f_K(\chi)$, where $K$ is the curvature parameter, and angular coordinates, $\pmb{\theta}$, is described by the matter density contrast
\be
    \delta(\pmb{x},\chi) = \frac{\rho(\pmb{x},\chi) - \Bar{\rho}(\chi)}{\Bar{\rho}(\chi)}\;,
\ee
where $\rho(\pmb{x},\chi)$ denotes the matter density with the spatial mean $\Bar{\rho}(\chi)$. 
The effective convergence can then be expressed as a weighted projection of the density contrast through 
\be
\label{eq:eff_convergence}
\kappa(\pmb{\theta}) = \frac{3H_0^2 \Omm}{2c^2} \int_0^{\chi_\mathrm{h}}\dd\chi~
    g(\chi)~f_K(\chi)~\frac{\delta(\pmb{x},\chi)}{a(\chi)}\;,
\ee
where we denote the matter density parameter by $\Omm$, the Hubble constant by $H_0$, the scale factor by $a$, the comoving horizon distance by $\chi_\mathrm{h}$ and define the weighting function 
\be
g(\chi') = \int_{z(\chi')}^{z(\chi_\mathrm{h})}\dd z~ p_{z}(z)~\frac{f_K(\chi(z)-\chi')}{f_K(\chi(z))}\;,
\ee
with the galaxy redshift distribution $p_z$. 

The discrete galaxy number density $n(\pmb{x},\chi)$ allows us to define the galaxy number density contrast as 
\be
    \delta_\mathrm{g}(\pmb{x},\chi) = \frac{n(\pmb{x},\chi) - \Bar{n}(\chi)}{\Bar{n}(\chi)}\;,
\ee
where $\Bar{n}(\chi)$ denotes the mean galaxy number density.
Similarly to Eq. \eqref{eq:eff_convergence}, we can define the projected galaxy number density contrast by 
\be
\kappa_\mathrm{g}(\pmb{\theta}) = \frac{N(\pmb{\theta}) - \Bar{N}}{\Bar{N}} 
    = \int\dd\chi~p_\mathrm{f}(\chi)~\delta_\mathrm{g}(\pmb{x},\chi)\;,
\ee
where $p_\mathrm{f}(\chi)$ describes the distribution of the foreground lens galaxies with projected galaxy number density $N(\pmb{\theta})$ which has the mean value $\Bar{N}$.

\subsection{Second- and third-order statistics}
In GGL one studies the correlation between matter and galaxy distribution.
Using the Fourier transforms of the matter and galaxy number density contrasts, $\hat{\delta}(\pmb{k},\chi)$ and $\hat{\delta}_\mathrm{g}(\pmb{k},\chi)$,
the cross-power spectrum is defined as the correlator 
\be
\langle\hat{\delta}(\pmb{k},\chi)~\hat{\delta}_\mathrm{g}(\pmb{k}',\chi)\rangle 
    = (2\pi)^3 \delta_\mathrm{D}(\pmb{k}+\pmb{k}')~P_{\delta\mathrm{g}}(|\pmb{k}|,\chi)\;.
\ee
The generally scale-dependent bias parameter, $b$, and the galaxy-mass correlation coefficient, $r$,
can be used to relate the cross-power spectrum, $P_{\delta\mathrm{g}}$, to the matter power spectrum, $P_{\delta\delta}$, via 
\be
P_{\delta\mathrm{g}}(|\pmb{k}|,\chi) = b(|\pmb{k}|,\chi)~r(|\pmb{k}|,\chi)~P_{\delta\delta}(|\pmb{k}|,\chi)\;.
\ee
One can define the projected version of the cross-power spectrum by
$\langle\hat{\kappa}(\pmb{\ell})~\hat{\kappa}_\mathrm{g}(\pmb{\ell}')\rangle = (2\pi)^2~\delta_\mathrm{D}(\pmb{\ell}+\pmb{\ell}')~P_{\kappa\mathrm{g}}(|\pmb{\ell}|)$,
which can then be expressed through 
\be
\label{eq:projected_power_spectrum}
P_{\kappa\mathrm{g}}(\ell) = \frac{3H_0^2 \Omm}{2c^2}\int\dd\chi~\frac{g(\chi)~p_\mathrm{f}(\chi)}{f_K(\chi)~a(\chi)} 
P_{\delta\mathrm{g}}\left(\frac{\ell}{f_K(\chi)},\chi\right)\;,
\ee
where one assumes a flat sky and uses Limber's equation \citep{Limber_1953,Kaiser_1998}.
From weak lensing surveys, the cross-power spectrum can be probed by the correlation function
\be
\label{eq:2ndorder_gscorr}
\langle\gamma_\mathrm{t}\rangle(\vartheta) 
    = \langle\kappa_\mathrm{g}(\pmb{\theta})~\gamma(\pmb{\theta}+\pmb{\vartheta};\varphi)\rangle,\;
\ee
where foreground galaxy positions are correlated with the rotated shear
\be
\gamma(\pmb{\theta};\varphi) = \gamma_\mathrm{t}(\pmb{\theta};\varphi) + \iunit \gamma_\times(\pmb{\theta};\varphi)
= -\gamma_\mathrm{c}(\pmb{\theta})~\eexp^{-2\iunit\varphi}\;,
\ee
which is the projection of the shear, $\gamma_\mathrm{c}$, as measured in a cartesian reference frame, towards the direction $\varphi$, being here the polar angle of the separation vector $\pmb{\vartheta}$.
The real and imaginary part of the rotated shear can be identified with the tangential and cross components. 

Similarly to Eq. \eqref{eq:2ndorder_gscorr}, we can also define the three-point correlation function (3PCF) according to \cite{Schneider_2005},
\be
\label{eq:3rdorder_ggscorr}
G_{\mathrm{gg}\gamma}(\vartheta_1,\vartheta_2,\phi) = \left\langle\kappa_\mathrm{g}(\pmb{\theta}+\pmb{\vartheta}_1)~\kappa_\mathrm{g}(\pmb{\theta}+\pmb{\vartheta}_2)~\gamma\left(\pmb{\theta};\frac{\varphi_1+\varphi_2}{2}\right)\right\rangle,
\ee
where $\phi$ is the angle between the vectors $\pmb{\vartheta}_1$ and $\pmb{\vartheta}_2$, and the rotated shear is projected in the direction of the bisecting line between the two foreground galaxies. 
This 3PCF also extends the notion of the average tangential shear $\langle\gamma_\mathrm{t}\rangle$ to the third order and is sensitive to the excess shear measured around pairs of foreground galaxies.

\subsection{\label{ssc:apstats_intro} Aperture measures}
A useful statistical measure to study the information contained in an observed shear field is provided by the aperture mass, developed by \cite{Schneider_1996}, defined by 
\be
\Map(\pmb{\vartheta}_0;\theta) = \int_{\mathbb{R}^2}\dd^2\vartheta~U_\theta(|\pmb{\vartheta}-\pmb{\vartheta}_0|)~\kappa(\pmb{\vartheta})\;,
\ee
which is the convolution of the convergence, $\kappa$, with a compensated filter function, $U_\theta$, within a circular aperture centered at angular position, $\pmb{\vartheta}_0$. 
Recalling that shear and convergence emerge from the same scalar potential, the aperture mass can also be expressed by a convolution of the tangential shear field with a different filter function, $Q_\theta$.
We define the complex aperture measure 
\be
\begin{alignedat}{1}
M = \Map + \iunit M_\times &= \int_{\mathbb{R}^2}\dd^2\vartheta~Q_\theta(|\pmb{\vartheta}-\pmb{\vartheta}_0|)~
[ \gamma_\mathrm{t}(\pmb{\vartheta};\pmb{\vartheta}_0) + \iunit\gamma_\times(\pmb{\vartheta};\pmb{\vartheta}_0) ] \\
&= \int_{\mathbb{R}^2}\dd^2\vartheta~Q_\theta(|\pmb{\vartheta}-\pmb{\vartheta}_0|)~\gamma(\pmb{\vartheta};\pmb{\vartheta}_0)\;,
\end{alignedat}
\ee
where, provided that the filter function $U_\theta$ is compensated, i.e. $\int\dd\vartheta~\vartheta~U_\theta(\vartheta) = 0$, 
the filter function $Q_\theta$ is given by
\be
Q_\theta(\vartheta) = \frac{2}{\vartheta^2}\int_0^\vartheta\dd\vartheta'~\vartheta' U_\theta(\vartheta') - U_\theta(\vartheta)\;.
\ee
The aperture scale, $\theta$, introduces a characteristic angular scale around which the filter functions contain most of their weight.
Since the quantity $M_\times$ vanishes if the shear field is solely due to lensing \citep{Crittenden_2002, Schneider_2002}, it is a measure of the $B$-mode contamination of the shear field. 
In turn, $\Map$ is only sensitive to $E$-mode shear such that the conversion of the shear field into the complex aperture measure $\Map + \iunit M_\times$ provides a discriminator of $E$- and $B$-mode components. 

In addition to the aperture mass, we define the aperture number counts, $\Nap$, \citep{Schneider_1998} by
\be
\begin{alignedat}{1}
\Nap(\pmb{\vartheta}_0;\theta) &= \int_{\mathbb{R}^2}\dd^2\vartheta~U_\theta(|\pmb{\vartheta}-\pmb{\vartheta}_0|)~
\kappa_\mathrm{g}(\pmb{\vartheta}) \\
&= \frac{1}{\Bar{N}}\int_{\mathbb{R}^2}\dd^2\vartheta~U_\theta(|\pmb{\vartheta}-\pmb{\vartheta}_0|)~N(\pmb{\vartheta})\;,
\end{alignedat}
\ee
where the second equality follows from $U_\theta$ being compensated. 
In analogy to the aperture mass, the aperture number counts vanish in the case of a constant galaxy number density $N$. 
This property ensures that $\Nap$ is only sensitive to deviations from the mean galaxy number density, and thus provides a smoothed representation of the galaxy number density field with zero mean. 
Throughout this work, we will use the filter functions
\be
\label{eq:filter_functions_definition}
\begin{alignedat}{1}
    U_\theta(\vartheta) &= \frac{1}{2\pi\theta^2}\left(1-\frac{\vartheta^2}{2\theta^2}\right)
    \exp\left(-\frac{\vartheta^2}{2\theta^2}\right)\;,\quad \text{and} \\
    Q_\theta(\vartheta) &= \frac{\vartheta^2}{4\pi\theta^4} \exp\left(-\frac{\vartheta^2}{2\theta^2}\right)\;, 
\end{alignedat}
\ee
first proposed by \cite{Crittenden_2002}.

\section{\label{sc:fourth-order_theory} Fourth-order statistics}
The concepts presented in the previous section are generalized to fourth-order statistical measures between shear and galaxies, in the following. 
Here we focus on the discussion of a single shear value correlated with the position of a triplet 
of foreground galaxies that are biased tracers of the underlying mass distribution. For an overview of the 
remaining fourth-order galaxy-shear correlation functions, see Appendix \ref{app:A_4PCFs}. 

\subsection{\label{ssc:trispectrum} Trispectrum}
The definition of the power spectrum via a second-order correlation function between the Fourier transforms 
of the matter density and galaxy number density contrast, $\hat{\delta}$ and $\hat{\delta}_\mathrm{g}$, can simply be generalized 
to the fourth order. The galaxy-galaxy-galaxy-mass trispectrum is defined by the correlator 
\be
\label{eq:gggmtrispectrum}
\begin{alignedat}{1}
    \langle\hat{\delta}_\mathrm{g}(\pmb{k}_1)&~\hat{\delta}_\mathrm{g}(\pmb{k}_2)~
    \hat{\delta}_\mathrm{g}(\pmb{k}_3)~\hat{\delta}(\pmb{k}_4)\rangle \\
    &= (2\pi)^3~\delta_\mathrm{D}(\pmb{k}_1+\pmb{k}_2+\pmb{k}_3+\pmb{k}_4)~T_{\mathrm{ggg}\delta}(\pmb{k}_1,\pmb{k}_2,\pmb{k}_3;\pmb{k}_4;\chi)\;.
\end{alignedat}
\ee
Due to statistical homogeneity of the matter and galaxy distributions, this correlator vanishes unless the four $\pmb{k}$-vectors form
a closed tetragon, which is ensured by the delta function. 
Moreover, statistical isotropy implies that $T_{\mathrm{ggg}\delta}$ depends only on the length of three of the $\pmb{k}$-vectors and two
angles enclosed by them.
In Eq. \eqref{eq:gggmtrispectrum}, we adopt the notation from \cite{Schneider_2005} and separate the arguments of the trispectrum by ``;'' in order to stress that the
corresponding function is invariant under permutations of its arguments left and right of ``;''. 
In this case any permutations of the arguments $\pmb{k}_1$, $\pmb{k}_2$, and $\pmb{k}_3$ will leave 
$T_{\mathrm{ggg}\delta}(\pmb{k}_1,\pmb{k}_2,\pmb{k}_3;\pmb{k}_4;\chi)$ invariant. 

Since galaxies are biased tracers of the matter distribution, $T_{\mathrm{ggg}\delta}$ can be expressed solely by the matter trispectrum, $T_{\delta\delta\delta\delta}$, defined by 
\be
\label{eq:mmmmtrispectrum}
\begin{alignedat}{1}
    \langle\hat{\delta}(\pmb{k}_1)&~\hat{\delta}(\pmb{k}_2)~\hat{\delta}(\pmb{k}_3)~\hat{\delta}(\pmb{k}_4)\rangle \\
    &= (2\pi)^3~\delta_\mathrm{D}(\pmb{k}_1+\pmb{k}_2+\pmb{k}_3+\pmb{k}_4)~
    T_{\delta\delta\delta\delta}(\pmb{k}_1,\pmb{k}_2,\pmb{k}_3,\pmb{k}_4;\chi)\;.
\end{alignedat}
\ee
By introducing the fourth-order bias parameter, $b_4$, and the galaxy-mass correlation coefficient, $r_4$, which are generally dependent on the length scale and comoving distance, this interrelation reads
\be
\label{eq:galaxybias_masscorrelationcoeff}
\begin{alignedat}{1}
    T_{\mathrm{ggg}\delta}(\pmb{k}_1,\pmb{k}_2,\pmb{k}_3;\pmb{k}_4;\chi) &= b_4^3~r_4~T_{\delta\delta\delta\delta}(\pmb{k}_1,\pmb{k}_2,\pmb{k}_3;\pmb{k}_4;\chi)\;.
\end{alignedat}
\ee
These two parameters describe the coupling between galaxies and the overall matter distribution in the Universe. 
Consequently, they depend on galaxy formation and evolution, and encode non-Gaussian information on the galaxy bias. 
In the case of linear deterministic biasing we expect these parameters to be $b_4=b$ and $r_4=1$, with $b$ being the bias factor relating $\delta_\mathrm{g}=b\delta$. 

The trispectrum as defined above describes the three-dimensional distribution of matter and galaxies with the third dimension being encoded by the comoving distance $\chi(z)$, a function of the redshift $z$. However, cosmic shear, as we observe it, rather emerges directly from the projection of the trispectrum along the line-of-sight. 
We denote the projected trispectrum by $t_{\mathrm{ggg}\kappa}$ and define it as 
\be
\label{eq:projected_trispectrum_def}
\begin{alignedat}{1}
    \langle\hat{\kappa}_\mathrm{g}(\pmb{\ell}_1)&~\hat{\kappa}_\mathrm{g}
    (\pmb{\ell}_2)~\hat{\kappa}_\mathrm{g}(\pmb{\ell}_3)~\hat{\kappa}(\pmb{\ell}_4)\rangle \\
    &= (2\pi)^2~\delta_\mathrm{D}(\pmb{\ell}_1+\pmb{\ell}_2+\pmb{\ell}_3+\pmb{\ell}_4)~
    t_{\mathrm{ggg}\kappa}(\pmb{\ell}_1,\pmb{\ell}_2,\pmb{\ell}_3;\pmb{\ell}_4)\;.
\end{alignedat}
\ee
This can be expressed through the matter trispectrum, $T_{\delta\delta\delta\delta}$, by integrating over the line-of-sight 
\be
\label{eq:trispectrum_projection}
\begin{alignedat}{1}
    t_{\mathrm{ggg}\kappa}&(\pmb{\ell}_1,\pmb{\ell}_2,\pmb{\ell}_3;\pmb{\ell}_4) \\
    &= \frac{3 H_0^2 \Omm}{2 c^2} \int \dd\chi~ \frac{g(\chi)~p_\mathrm{f}^3(\chi)}{f_K^5(\chi)~a(\chi)}~
    T_{\mathrm{ggg}\delta} \left(\pmb{k}_1, \pmb{k}_2, \pmb{k}_3; \pmb{k}_4; \chi \right)\,,
\end{alignedat}
\ee
where we defined the vectors $\pmb{k}_i = \pmb{\ell}_i/f_K(\chi)$. 
The derivation of the above equation is analogous to the one in order to obtain the projected power spectrum in Eq. \eqref{eq:projected_power_spectrum}. 
This equation shows that the observed correlations between galaxies and matter depend on one hand on the redshift distributions of the observed galaxy populations, but also on the galaxy bias and galaxy-mass correlation coefficient, as implied by by Eq. \eqref{eq:galaxybias_masscorrelationcoeff}. 

\subsection{\label{ssc:4PCF} Four-point correlation function}
\begin{figure}
    \centering
    \usetikzlibrary {positioning}
\usetikzlibrary{arrows}
\begin{tikzpicture}
    \pgfmathsetmacro{\xone}{cos(\firstdir)}%
    \pgfmathsetmacro{\yone}{sin(\firstdir)}%
    \pgfmathsetmacro{\xtwo}{cos(\seconddir)}%
    \pgfmathsetmacro{\ytwo}{sin(\seconddir)}%
    \pgfmathsetmacro{\xthree}{cos(\thirddir)}%
    \pgfmathsetmacro{\ythree}{sin(\thirddir)}%
    \coordinate (a) at ($\length*(\xone,\yone)$);
    \coordinate (b) at ($\length*(\xtwo,\ytwo)$);
    \coordinate (c) at ($\length*(\xthree,\ythree)$);
    \coordinate (o) at (0,0);
    
    \draw[line width=1pt] [->] (0,0) -- (a) node [below,pos=0.7] {\Large$\pmb{\vartheta}_1$};
    \draw[line width=1pt] [->] (0,0) -- (b) node [right,pos=0.8] {\Large$\pmb{\vartheta}_2$};
    \draw[line width=1pt] [->] (0,0) -- (c) node [left,pos=0.5] {\Large$\pmb{\vartheta}_3$};

    \draw[line width=2pt] (-0.3,0) -- (0.3,0) node [below,pos=0] {\Large$\pmb{\theta}_4$};
    \filldraw[black] (a) circle (3pt) node (x) {};
    \node[below=24pt, label=\Large$\pmb{\theta}_1$] at (a) {};
    \filldraw[black] (b) circle (3pt) node (x) {};
    \node[below left=15pt, label=\Large$\pmb{\theta}_2$] at (b) {};
    \filldraw[black] (c) circle (3pt) node (x) {};
    \node[below left=15pt, label=\Large$\pmb{\theta}_3$] at (c) {};
    
    \draw [->] ($2*(\xone,\yone)$)  arc (\firstdir:\seconddir:2) node [above,pos=0.5] {\Large$\phi_{12}$};
    \draw [->] ($2*(\xtwo,\ytwo)$)  arc (\seconddir:\thirddir:2) node [above,pos=0.5] {\Large$\phi_{23}$};
    \draw [->] ($3*(\xone,\yone)$)  arc (\firstdir:\thirddir:3) node [above left,pos=0.5] {\Large$\phi_{13}$};

    \draw [dashed] (a) -- ($1.2*\length*(\xone,\yone)$);
    \draw [dashed] (a) -- ($\length*(\xone,\yone)+(0.2*\length,0)$);
    \draw [dashed] (b) -- ($1.2*\length*(\xtwo,\ytwo)$);
    \draw [dashed] (b) -- ($\length*(\xtwo,\ytwo)+(0.2*\length,0)$);
    \draw [dashed] (c) -- ($1.2*\length*(\xthree,\ythree)$);
    \draw [dashed] (c) -- ($\length*(\xthree,\ythree)+(0.2*\length,0)$);

    \draw [->] ($\length*(\xone,\yone)+(0.1*\length,0)$)  arc (0:\firstdir:0.45) node [right,pos=0.5] {\Large$\varphi_1$};
    \draw [->] ($\length*(\xtwo,\ytwo)+(0.1*\length,0)$)  arc (0:\seconddir:0.45) node [above right,pos=0.5] {\Large$\varphi_2$};
    \draw [->] ($\length*(\xthree,\ythree)+(0.1*\length,0)$)  arc (0:\thirddir:0.45) node [above right,pos=0.5] {\Large$\varphi_3$};

\end{tikzpicture}
    \caption{Sketch of a galaxy-galaxy-galaxy-shear configuration which the 4PCF $\nnngCF$ is sensitive to.  The configuration is fully characterized by the radial separations $\vartheta_1$, $\vartheta_2$ and $\vartheta_3$ and two of the three opening angles $\phi_{12}$, $\phi_{13}$ and $\phi_{23}$ which are related by $\phi_{13}=\phi_{12}+\phi_{23}$.}
    \label{fig:gggs_sketch}
\end{figure}
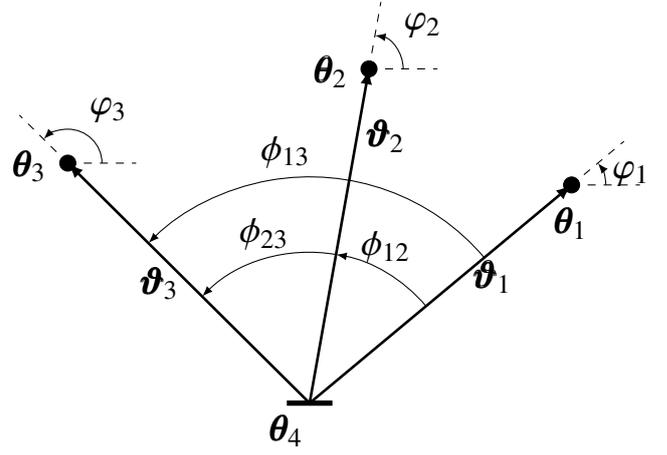

The galaxy-galaxy-galaxy-shear correlation function is sensitive to galaxy quadruplet configurations as depicted in Fig. \ref{fig:gggs_sketch}, where the shear at the position of a background galaxy, $\pmb{\theta}_4$, is measured around a triplet of foreground galaxies located at $\pmb{\theta}_1,\pmb{\theta}_2$ and $\pmb{\theta}_3$. 
We define this four-point correlation function (4PCF) as 
\be
\label{eq:Ggggk_definition}
\begin{alignedat}{1}
    \nnngCF(\pmb{\vartheta}_1,\pmb{\vartheta}_2,\pmb{\vartheta}_3) 
    &= \nnngCF(\vartheta_1,\vartheta_2,\vartheta_3,\phi_{12},\phi_{23}) \\
    &= \langle\kappa_\mathrm{g}(\pmb{\theta}_1)~\kappa_\mathrm{g}(\pmb{\theta}_2)~
    \kappa_\mathrm{g}(\pmb{\theta}_3)~\gamma(\pmb{\theta}_4;\zeta)\rangle\;,
\end{alignedat}
\ee
where the $\pmb{\vartheta}_i = \pmb{\theta}_i-\pmb{\theta}_4$, with lengths $\vartheta_i = |\pmb{\vartheta}_i|$, denote the angular separation vectors of each foreground galaxy to the background galaxy at position $\pmb{\theta}_4$. 
In this definition of the 4PCF, we apply a general projection angle $\zeta$ to the rotated shear, whose properties we shall discuss below. 
The opening angles $\phi_{12}$, $\phi_{23}$ and $\phi_{13}$ in the galaxy quadruplet are simply related by $\phi_{13}=\phi_{12}+\phi_{23}$, meaning the correlator depends only on two of the three opening angles. 
A practical estimator can only be obtained for the modified correlator 
\be
\label{eq:modifiedGgggk_definition}
\begin{alignedat}{1}
    \tilde{G}_{\mathrm{ggg}\gamma}(\vartheta_1,\vartheta_2,\vartheta_3,\phi_{12},\phi_{23}) 
    = \frac{1}{\Bar{N}^3} \langle N(\pmb{\theta}_1)~N(\pmb{\theta}_2)~N(\pmb{\theta}_3)~\gamma(\pmb{\theta}_4;\zeta)\rangle\;,
\end{alignedat}
\ee
that also contains information about lower-order correlators. Using the fact that $N/\Bar{N}=1+\kappa_\mathrm{g}$ and that the direction of projection of the rotated shear can be transformed according to $\gamma(\pmb{\theta};\phi) = \gamma(\pmb{\theta};\phi')~\eexp^{2\iunit(\phi'-\phi)}$, we obtain the relation 
\begin{equation}
\label{eq:modifiedGgggk_Ggggk_def}
\begin{alignedat}{1}
    \tilde{G}&_{\mathrm{ggg}\gamma}(\vartheta_1,\vartheta_2,\vartheta_3,\phi_{12},\phi_{23}) 
    = \nnngCF(\vartheta_1,\vartheta_2,\vartheta_3,\phi_{12},\phi_{23}) \\
    &\quad + \Bigg[\langle\gamma_\mathrm{t}\rangle(\vartheta_1)~\eexp^{2\iunit\varphi_1} 
    + \langle\gamma_\mathrm{t}\rangle(\vartheta_2)~\eexp^{2\iunit\varphi_2} 
    + \langle\gamma_\mathrm{t}\rangle(\vartheta_3)~\eexp^{2\iunit\varphi_3} \\
    &\qquad + G_{\mathrm{gg}\gamma}(\vartheta_1,\vartheta_2,\phi_{12})~\eexp^{\iunit(\varphi_1+\varphi_2)} 
    + G_{\mathrm{gg}\gamma}(\vartheta_2,\vartheta_3,\phi_{23})~\eexp^{\iunit(\varphi_2+\varphi_3)} \\
    &\qquad + G_{\mathrm{gg}\gamma}(\vartheta_1,\vartheta_3,\phi_{12}+\phi_{23})~\eexp^{\iunit(\varphi_1+\varphi_3)} \Bigg]
    ~\eexp^{-2\iunit\zeta}\;,
\end{alignedat}
\end{equation}
where we note that the shear projection angles in the 2PCFs and 3PCFs are chosen in accordance with Eqs. \eqref{eq:2ndorder_gscorr} and \eqref{eq:3rdorder_ggscorr}. 
Since the correlator $\tilde{G}_{\mathrm{ggg}\gamma}$ should not depend on the individual angles 
$\varphi_1,\varphi_2,\varphi_3$ but rather on $\phi_{12}$ and $\phi_{23}$, we have to make an appropriate 
choice for the projection angle $\zeta$. For that, we write it in the general form   
$\zeta = a_1\varphi_1 + a_2\varphi_2 + a_3\varphi_3$, i.e. as a linear combination of the polar angles of 
the positions of the foreground galaxies. 
Now, any valid choice for $\zeta$ has to meet the two following important properties: 
\begin{enumerate}
    \item For one, the projected shear $\gamma(\pmb{\theta}_4;\zeta) = -\gamma_\mathrm{c}(\pmb{\theta}_4)~\eexp^{-2\iunit\zeta}$ has to stay invariant under the transformation $\varphi_i \rightarrow \varphi_i + 2\pi$. 
    The position of one foreground galaxy rotated by $2\pi$ will thus result in the same rotated shear as for the non-rotated position. From this follows that the coefficients $a_i$ can only have discrete values $a_i \in \frac{1}{2} \mathbb{Z}$.
    \item Additionally, the projected shear has to stay invariant under a rotation of the whole coordinate frame by an arbitrary angle $\alpha$. The cartesian shear transforms like $\gamma_\mathrm{c} \rightarrow \gamma_\mathrm{c}~\eexp^{-2\iunit\alpha}$ and thus $\zeta$ has to transform like $\zeta \rightarrow \zeta - \alpha$ in order for the projected shear to stay invariant.
    The polar angles transform like $\varphi_i \rightarrow \varphi_i - \alpha$ and therefore the coefficients need to satisfy the identity $a_1 +a_2 + a_3 = 1$.
\end{enumerate}
The choice for $\zeta$ is arbitrary, as long as it meets the two conditions above.
Choosing another projection angle $\zeta'=a_1'\varphi_1 + a_2'\varphi_2 + a_3'\varphi_3$ changes the 4PCF only by the phase factor 
$\eexp^{2\iunit(\zeta'-\zeta)} = \eexp^{2\iunit[(a_1-a_1')\phi_{12} + (a_3'-a_3)\phi_{23}]}$.
Throughout the following work, we choose the shear projection angle to be $\zeta = (\varphi_1 + \varphi_3)/2$.
The definitions of the remaining galaxy-shear 4PCFs with multiple shear components are introduced in Sect. \ref{app:A_4PCFs}.

A measurement of $\nnngCF$ contains information about the corresponding trispectrum, $t_{\mathrm{ggg}\kappa}$, and thus can be used to study its properties. 
Similar to the 4PCF, the trispectrum depends only on the length of three of the vectors $\pmb{\ell}_i$, denoted by $\ell_i$, and two angles $\psi_{12}$ and $\psi_{23}$ enclosed by them. 
In principle, one can convert one quantity into the other by computing the integrals over the respective parameter space that read 
\be
\label{eq:4PCF_into_trispectrum}
\begin{alignedat}{1}
    \nnngCF&(\vartheta_1,\vartheta_2,\vartheta_3,\phi_{12},\phi_{23}) \\
    &= \left(\prod\limits_{i=1}^3 \int_0^\infty \frac{\dd\ell_i~\ell_i}{2\pi}\right) \int_0^{2\pi}\frac{\dd\psi_{12}}{2\pi}\int_0^{2\pi}\frac{\dd\psi_{23}}{2\pi}~
    t_{\mathrm{ggg}\kappa}~f_{\mathrm{ggg}\kappa}\;,
\end{alignedat}
\ee
and
\be
\label{eq:trispectrum_into_4PCF}
\begin{alignedat}{1}
    t_{\mathrm{ggg}\kappa}&(\ell_1,\ell_2,\ell_3,\psi_{12},\psi_{23}) \\
    &= 2\pi \left(\prod\limits_{i=1}^3 \int_0^\infty \dd\vartheta_i~\vartheta_i\right)\int_0^{2\pi}\dd\phi_{12}\int_0^{2\pi}\dd\phi_{23}~
    \nnngCF~f^*_{\mathrm{ggg}\kappa}\;,
\end{alignedat}
\ee
where we dropped the dependencies of the integrated functions. The filter function, $f_{\mathrm{ggg}\kappa}$, depends on the paramater space of both the 4PCF and the trispectrum and contains a second-order Bessel function. For a more detailed derivation of these relations and explicit expressions of the filter function, see Sect. \ref{app:A_4PCF_trispectra}. 

The numerical evaluation of the relations between the 4PCF and the trispectrum is in practice not trivial to realize, making it challenging to compare a measured 4PCF to a theoretical prediction obtained from a modelled trispectrum. 
In order to avoid this problem, we will consider the conversion of the 4PCF and the trispectrum to aperture statistics, in the following. 

\subsection{\label{ssc:fourth-order_apstats} G4L aperture statistics}
Here, we focus on the galaxy-galaxy-galaxy-mass aperture statistics $\langle\mathcal{N}^3 M\rangle$ which contain the compressed information content of the 4PCF $\nnngCF$.
These aperture statistics generally depend on four aperture scales, $\theta_1,\theta_2,\theta_3$ and $\theta_4$, and are given by the five-dimensional integral
\be
\label{eq:N3M_final}
\begin{alignedat}{1}
    \langle\mathcal{N}^3 M\rangle&(\theta_1,\theta_2,\theta_3;\theta_4) \\
    &= \left(\prod\limits_{i=1}^3 \int_0^\infty \dd\vartheta_i~\vartheta_i\right) \int_0^{2\pi}\dd\phi_{12} \int_0^{2\pi}\dd\phi_{23}~
    \nnngCF~A_{\mathcal{N}^3M}\;,
\end{alignedat}
\ee
where the filter function $A_{\mathcal{N}^3M}$ also depends on the parameter space of the 4PCF and all four aperture scales.  
This formula is derived in Sect. \ref{app:A_apstats} where we also explicitly derive the filter function $A_{\mathcal{N}^3M}$, which is plotted for some exemplary configurations in Fig. \ref{fig:AN3M_pcolor}. 
Moreover, the aperture statistics can also be expressed in terms of the projected trispectrum 
\be
\label{eq:N3M_tgggk}
\begin{alignedat}{1}
    \langle\Nap^3\Map\rangle&(\theta_1,\theta_2,\theta_3;\theta_4) 
    = \left(\prod\limits_{i=1}^3 \int_{\mathbb{R}^2} \frac{\dd^2\ell_i}{(2\pi)^2}~\hat{u}(\ell_i\theta_i)\right) \\
    \times& \hat{u}(|\pmb{\ell}_1+\pmb{\ell}_2+\pmb{\ell}_3|\theta_4) 
    \; t_{\mathrm{ggg}\kappa}(\pmb{\ell}_1,\pmb{\ell}_2,\pmb{\ell}_3;-\pmb{\ell}_1-\pmb{\ell}_2-\pmb{\ell}_3)\;,
\end{alignedat}
\ee
where we defined $U_{\theta}(\vartheta) = \theta^{-2}u(\vartheta/\theta)$ and used the Fourier transform $\hat{u}(\eta) = \int\dd^2x~u(|\pmb{x}|)~\eexp^{\iunit\pmb{x}\cdot\pmb{\eta}} = \frac{\eta^2}{2}\eexp^{-\eta^2/2}$ \citep{Schneider_2005}, assuming the filter function in Eq. \eqref{eq:filter_functions_definition}. 
The fact that $\hat{u}$ has a narrow peak indicates that the aperture statistics also directly reflect the local properties of the trispectrum. 
It is therefore evident that the aperture statistics are useful for studying the bias parameter and galaxy-mass correlation coefficient, which enter the equation through the trispectrum given in Eq. \eqref{eq:trispectrum_projection}.

The conversion of the 4PCF into aperture statistics has the great advantage that it heavily reduces the size of our data vector, from millions of bins in the 4PCF to a few hundreds or a few thousands, if we use multiple different aperture scales.
In contrast to using the direct relations \eqref{eq:4PCF_into_trispectrum} and \eqref{eq:trispectrum_into_4PCF} between trispectrum and 4PCF, aperture statistics require the 4PCF to be measured only in a limited range of the radial coordinates $\vartheta_1,\vartheta_2,\vartheta_3$ due to the filter functions $U_\theta$ and $Q_\theta$ being only significantly nonzero in a narrow domain. 
Moreover, in Eq. \eqref{eq:N3M_final} we can replace $\nnngCF$ by its modified version $\tilde{G}_{\mathrm{ggg}\gamma}$, Eq. \eqref{eq:modifiedGgggk_definition}, for which it is possible to construct a practical estimator. 
The $\langle\mathcal{N}^3 M\rangle$ statistics remain invariant under that replacement, making them blind to the signal from the lower-order correlators appearing in Eq. \eqref{eq:modifiedGgggk_Ggggk_def}. 
This highlights the fact that the aperture statistics constitute a direct measure of the full four-point contribution to the measured 4PCF signal. 
Moreover, we do not have to worry about including the lower-order corrections in the calculation of $\langle\mathcal{N}^3 M\rangle$ which would come along with additional numerical and systematical biases. 

\section{\label{sc:measureG4L} Measuring the G4L aperture statistics with a numerical integration pipeline}
With Eq. \eqref{eq:N3M_final} and the expression for the filter function in Eq. \eqref{eq:app_AN3M} at hand, we are now in the position to compute $\langle\Nap^3\Map\rangle$ once the 4PCF $\nnngCF$ is measured for a certain configuration of bins in the $(\vartheta_1,\vartheta_2,\vartheta_3,\phi_{12},\phi_{23})$-space. 
For a numerical integration pipeline based on a Riemann sum, the computational complexity is severely affected by the high dimensionality of that parameter space.
Therefore the bin configurations on which the 4PCF is evaluated faces constraints in its resolution; this affects mainly the three radial variables $\vartheta_1,\vartheta_2$ and $\vartheta_3$. 
In the end, the numerical integration needs to yield sufficiently accurate results under the constraint that the number of bins is strictly limited, while also running for a reasonable amount of time. 
In this section, we investigate the technical details of a numerical integration routine based on a Riemann sum and assess the precision of its results.

Before we apply the numerical integration on a complete 4PCF, we first verify the correctness of the results by only taking the Gaussian part of the 4PCF into account, meaning we neglect the connected part for the time being. 
This has the advantage that we can decompose the 4PCF into the respective 2PCFs, as shown in Eq. \eqref{eq:Ggggk_gaussian_fields}. 
The 2PCFs can be easily measured with high resolution, which enables us to analyze the performance of the numerical integration pipeline in terms of accuracy. 
In this special case, the results can be checked because -- like the 4PCF -- the fourth-order aperture statistics can also be decomposed into second-order aperture statistics (see App. \ref{app:numint}), which can be readily measured. 

To generate theory predictions for the 2PCFs $\langle\kappa_\mathrm{g}\kappa_\mathrm{g}\rangle(\vartheta)$ and $\langle\kappa_\mathrm{g}\gamma\rangle(\vartheta)$, we employ the CCL-code of \cite{Chisari_2019}. 
We assume a flat $\Lambda$CDM cosmology with the matter density parameter set to $\Omega_\mathrm{m}=0.3$, the normalized Hubble constant $h=0.7$, the amplitude of density fluctuations $\sigma_8=0.8$, and the spectral index $n_\mathrm{s}=0.96$, and choose a BBKS transfer function \citep{Bardeen_1986} to compute the linear power spectrum using the halofit prescription of \cite{Takahashi_2012} with a linear deterministic bias $b=1$. 
In these mocks, both the lens and the source galaxies follow a Gaussian distribution in redshift space with mean redshifts 1 and 1.5, respectively, and a standard deviation of 0.15. 

\subsection{\label{ssc:numint_method} Numerical integration method}
For the purpose of the numerical integration, we apply a dimensionless transformation of the radial integration variables to  $\ln y_i=\ln(\vartheta_i/\theta_i)$ in Eq. \eqref{eq:N3M_final}. 
We sample the logarithmic, dimensionless radial integration variable $\ln y_i$ in $N_y$ bins in the integration range $[\ln y_\mathrm{min},\ln y_\mathrm{max}]$, using a constant logarithmic bin size $\Dlog$. 
The angular variables, $\phi_{12}$ and $\phi_{23}$, are each linearly spaced in $N_\phi$ bins of size $\Delta_\phi$ in the domain $[-\pi,\pi]$. 
With this binning scheme, the estimator of the integral reads
\be
\label{eq:N3M_riemannsum}
\begin{alignedat}{1}
    \langle\mathcal{N}^3 M\rangle_\mathrm{est}&(\theta_1,\theta_2,\theta_3;\theta_4) 
    = \sum\limits_{i_1,i_2,i_3=1}^{N_y} \Dlog^3 ~ y_{i_1}^2 y_{i_2}^2 y_{i_3}^2 \theta_1^2\theta_2^2\theta_3^2 \\
    \times& \sum\limits_{j_1,j_2=1}^{N_\phi} \Delta_\phi^2 ~ \nnngCF( y_{i_1} \theta_1, y_{i_2} \theta_2, y_{i_3} \theta_3,\phi_{j_1},\phi_{j_2}) \\
    \times& ~A_{\mathcal{N}^3M}( y_{i_1} \theta_1, y_{i_2} \theta_2, y_{i_3} \theta_3,\phi_{j_1},\phi_{j_2}|\theta_1,\theta_2,\theta_3;\theta_4)\;,
\end{alignedat}
\ee
where we evaluate the 4PCF and the filter function at the centers of the respective bins. 
The main technical issue of this method is the high number of function evaluations, $N_y^3N_\phi^2$. 
The impact of the high computational complexity can be mitigated by minimizing the radial integration range and by adjusting the angular binning scheme -- all in compliance with high-precision results. 

\subsection{Effect of the integration range}
Although formally we need to evaluate the integration over the radial coordinates $\vartheta_i$ in Eq. \eqref{eq:N3M_final} from zero to infinity, in practice we have to restrict the integration range due to finite field sizes and resolution limits. 
Such a restriction will generally induce a bias in the final result of the integration.
From a computational perspective, it is desirable to reduce the integration range as much as possible, provided that the induced bias is small enough. 

\begin{figure}
    \centering
    \includegraphics[width=\columnwidth]{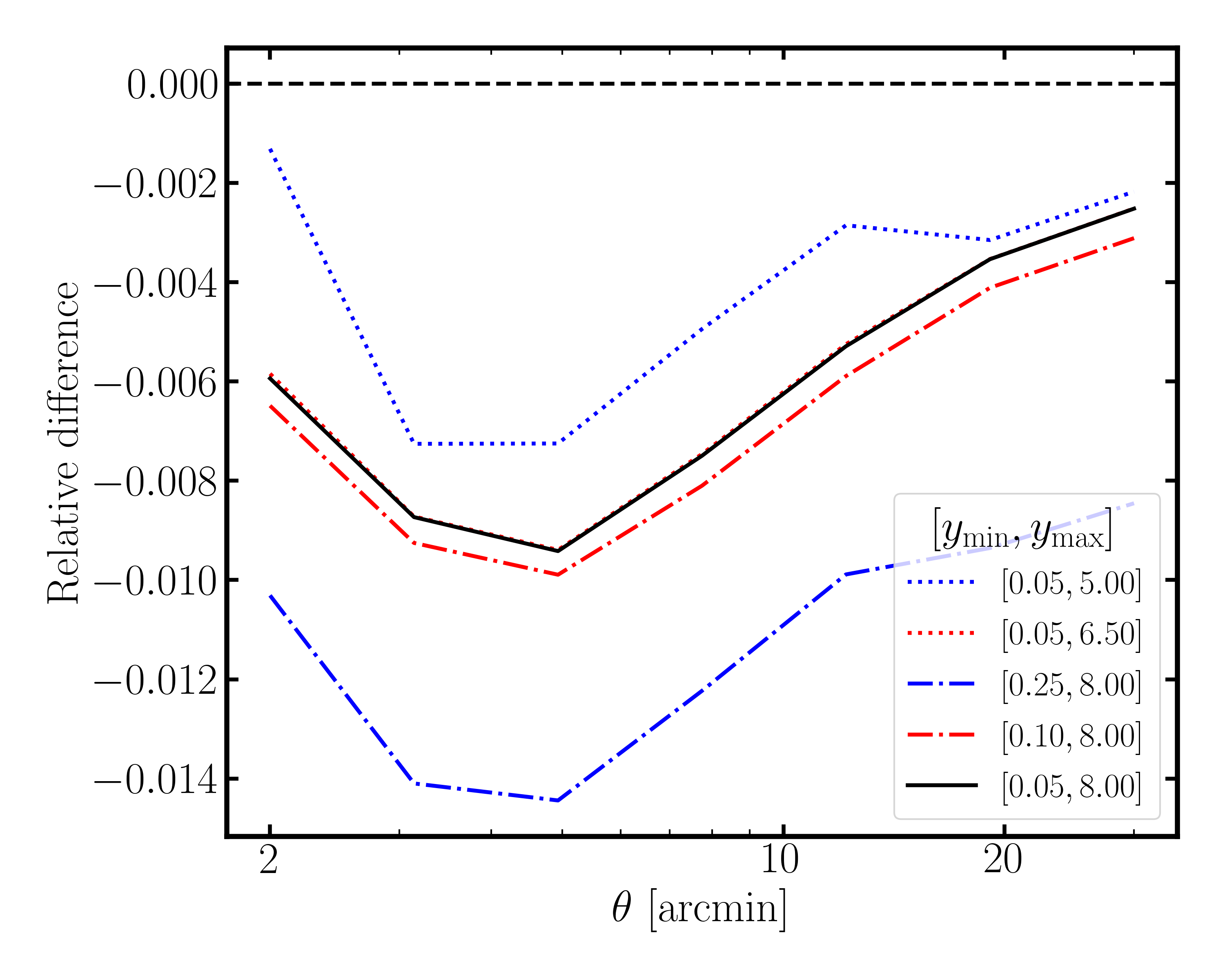}
    \caption{Relative difference of the numerical integration results for $\langle\Nap^3\Map\rangle$ according to Eq. \eqref{eq:N3M_riemannsum} with respect to the expected results obtained from the second-order aperture statistics. The integration was performed using a Gaussian 4PCF, and with different configurations of the integration range $[y_\mathrm{min},y_\mathrm{max}]$, all using $\Dlog \approx 0.11$ and $\Dphi=1/350$. The dotted (dash-dotted) lines show that the results converge with an increasing (decreasing) upper (lower) integration boundary.}
    \label{fig:numint_minmaxratio}
\end{figure}

In the following, we vary the lower and upper integration boundaries of the variables $y_i\in[y_\mathrm{min},y_\mathrm{max}]$ in Eq. \eqref{eq:N3M_riemannsum} to study their effect on the integration result.
In Fig. \ref{fig:numint_minmaxratio}, we show the relative difference of the numerical integration results with respect to the expected results obtained from the second-order aperture statistics\footnote{The second-order aperture statistics are obtained via numerical integration (see App. \ref{app:numint}), using an integration range of $[0\overset{\prime}{.}1,250']$.},
for a number of choices for $[y_\mathrm{min},y_\mathrm{max}]$ and equal aperture scales. 
The results converge to the curve with the integration range $[0.05,8.00]$ which will be used as the integration boundaries throughout this paper. 
With the minimum aperture scale of $2'$, the lower integration boundary implies that the 4PCF would need to be measured down to scales of $0\overset{\prime}{.}1$.  

We note that these integration boundaries are suitable for the case of equal aperture scales. 
Once we perform the numerical integration with different scales $\theta_1,\theta_2,\theta_3$ and $\theta_4$, the upper integration boundary is affected in particular and needs to be raised, leading to an increased computational complexity. 

\subsection{\label{ssc:angular_binning}Binning scheme of the angular part}
\begin{figure}[t]
    \centering
    \includegraphics[width=\columnwidth]{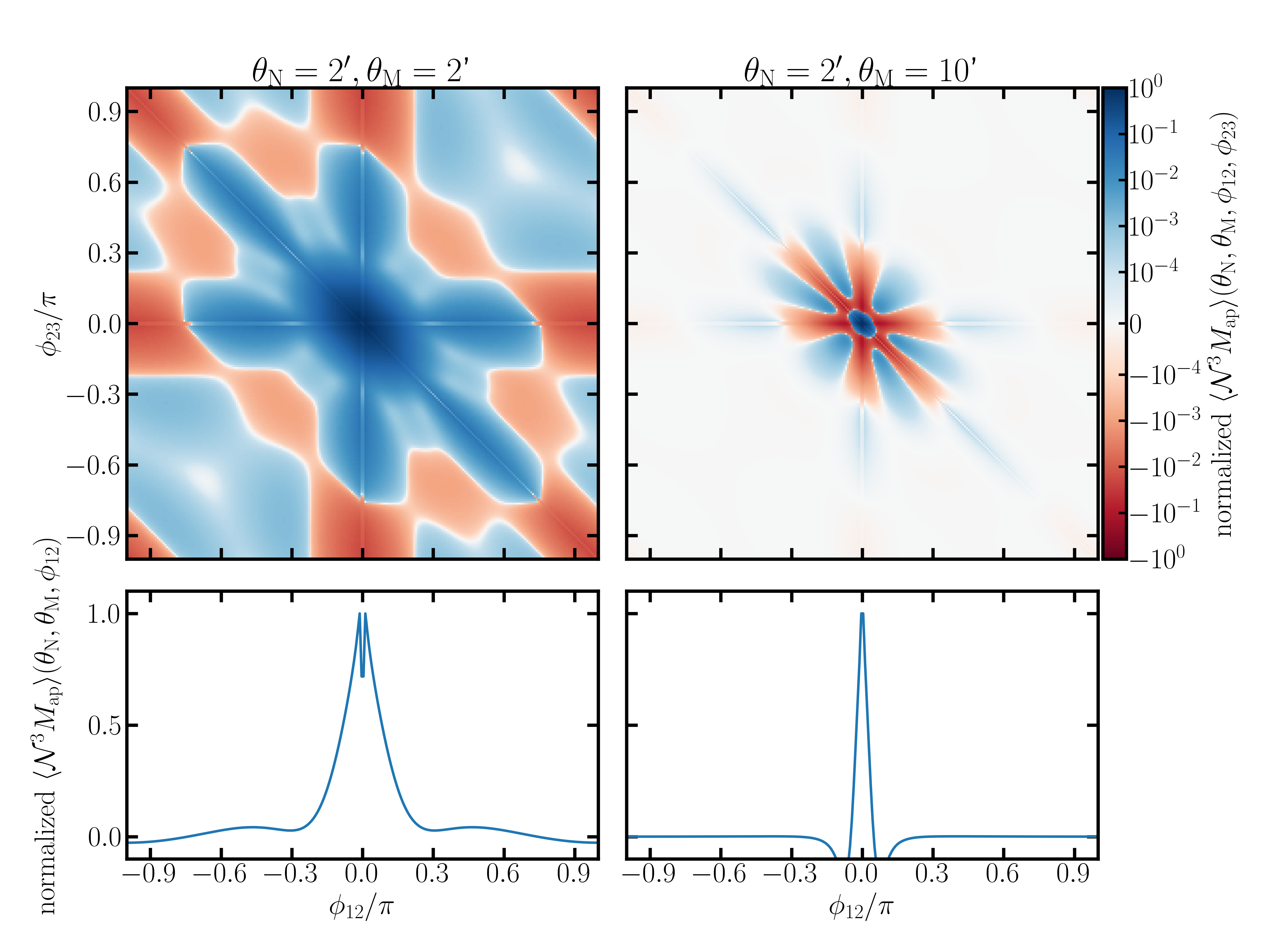}
    \caption{The real part of Eq. \eqref{eq:N3M_riemannsum} as a function of the angular variables $\phi_{12}$ and $\phi_{23}$, with the integrations over the three radial variables already performed. All results are normalized by the respective maximum value. 
    While the upper plots show the 2-D structure as a function of both $\phi_{12}$ and $\phi_{23}$, the lower plots show the results from the upper plots marginalized over the $\phi_{23}$-axis.
    The left and right panel show different configurations of the aperture scales, where we set $\thap{N}{}=\theta_1=\theta_2=\theta_3$ and $\thap{M}{}=\theta_4$.
    The values of the color scale are spaced logarithmically for both positive and negative values with a linear domain for values in the interval $[-10^{-4},10^{-4}]$. 
    The numerical integration was performed using $\Dlog \approx 0.16$ and $\Delta_\phi/2\pi= 1/250$.
    }
    \label{fig:integrand_angular_part}
\end{figure}

To reduce the computational complexity of the numerical integration, we consider an adaptive bin size in the integration domain of the angular part of Eq. \eqref{eq:N3M_riemannsum}, i.e. in the $(\phi_{12},\phi_{23})$-plane. 
For that, it is useful to inspect the integrand once the integration over all three radial variables has been performed. 
This is shown in Fig. \ref{fig:integrand_angular_part} where we can see that the angular part of the numerical integration does not show sharp features in a large part of the integration domain. 
For large values of both $\phi_{12}$ and $\phi_{23}$, a coarse bin size, $\Delta_\phi$, might already be sufficient in order to sample the integrand adequately. 

As can also be seen in Fig. \ref{fig:integrand_angular_part}, most of the weight of the integral lies around the central region where $\phi_{12}$ and $\phi_{23}$ are small; therefore, it is important to accurately sample the integrand in this region. 
In fact, we require a fine bin size, $\Delta_\phi$, in the integration domain where $\phi_{12}$, $\phi_{23}$ or $\phi_{12}+\phi_{23}$ are close to zero.

From here on, we apply an adaptive binning scheme in the $(\phi_{12},\phi_{23})$-plane, which uses a coarse bin size of $\Dphi=1/50$ and a sufficiently small bin size only in the regions where $\phi_{12}$, $\phi_{23}$ or $\phi_{12}+\phi_{23}$ are close to zero. 
This reduces the number of function evaluations in the angular part to a fraction of $N_\phi^2 = (2\pi/\Delta_\phi)^2$ with $\Delta_\phi$ being the small bin size.
By performing several test integrations, we find that, in order to obtain highly accurate results, it is sufficient to apply a finer binning with $\Dphi=1/350$ in a sub-domain of the $(\phi_{12},\phi_{23})$-plane where either $\phi_{12}$, $\phi_{23}$ or $\phi_{12}+\phi_{23}$ lie in the range $[-0.2\pi,0.2\pi]$.

\subsubsection{\label{sc:bin-avg_4PCF}Bin-averaged 4PCF}
In the measurement of a 4PCF, there is generally more than one galaxy quadruplet configuration that falls into one specific bin of the parameter space $(\vartheta_1,\vartheta_2,\vartheta_3,\phi_{12},\phi_{23})$. 
Although all configurations have very similar radial and angular coordinates, their parameters are slightly different from those defining the bin center. 
For this reason, the resulting measured 4PCF represents an average over all configurations that fall into each bin. 
Consequently, we need to replace the 4PCF in Eq. \eqref{eq:N3M_riemannsum} with the bin-averaged 4PCF according to 
\be
\begin{alignedat}{1}
\Bar{G}_{\mathrm{ggg}\gamma}&(\vartheta_1,\vartheta_2,\vartheta_3,\phi_{12},\phi_{23}) \\
=& \frac{1}{\Dlog^3} \left[\prod_{i=1}^3 \int_{\ln\vartheta_i-\Dlog/2}^{\ln\vartheta_i+\Dlog/2} \dd\ln\vartheta_i'~\right]~
\nnngCF(\vartheta_1',\vartheta_2',\vartheta_3',\phi_{12},\phi_{23})\;.
\end{alignedat}
\ee

In our case, we integrate only the Gaussian part of the 4PCF, which is calculated from 2PCFs according to Eq. \eqref{eq:Ggggk_gaussian_fields}.
In order to imitate the bin-averaged 4PCF, we also need to average the Gaussian 4PCF. 
This can be achieved by subdividing every integration bin into $N_\mathrm{avg}$ smaller sub-bins and averaging the Gaussian 4PCF over those sub-bins. 
We can compute this bin-averaged Gaussian 4PCF by 
\be
\begin{alignedat}{1}
\Bar{G}^\mathrm{Gauss}_{\mathrm{ggg}\gamma}&(\vartheta_1,\vartheta_2,\vartheta_3,\phi_{12},\phi_{23}) \\
&= \frac{1}{N_\mathrm{avg}^3}\sum_{i_1,i_2,i_3=-(N_\mathrm{avg}-1)/2}^{(N_\mathrm{avg}-1)/2} 
G^\mathrm{Gauss}_{\mathrm{ggg}\gamma}\left(\vartheta_1^\mathrm{sub}, \vartheta_2^\mathrm{sub},\vartheta_3^\mathrm{sub},\phi_{12},\phi_{23}\right)\,,
\end{alignedat}
\ee
where the $\vartheta_1,\vartheta_2,\vartheta_3$ are the centers of the integration bins, and the sub-bins are sampled symmetrically in log-space around these centers, located at $\vartheta_k^\mathrm{sub} = \vartheta_k\exp\left(i_k~\Dlog/N_\mathrm{avg}\right)$ with $k\in\{1,2,3\}$. 
The Gaussian 4PCF is then sampled with an effective logarithmic bin size $\Dlog/N_\mathrm{avg}$. 

\begin{figure}[htbp]
    \centering
    \includegraphics[width=\columnwidth]{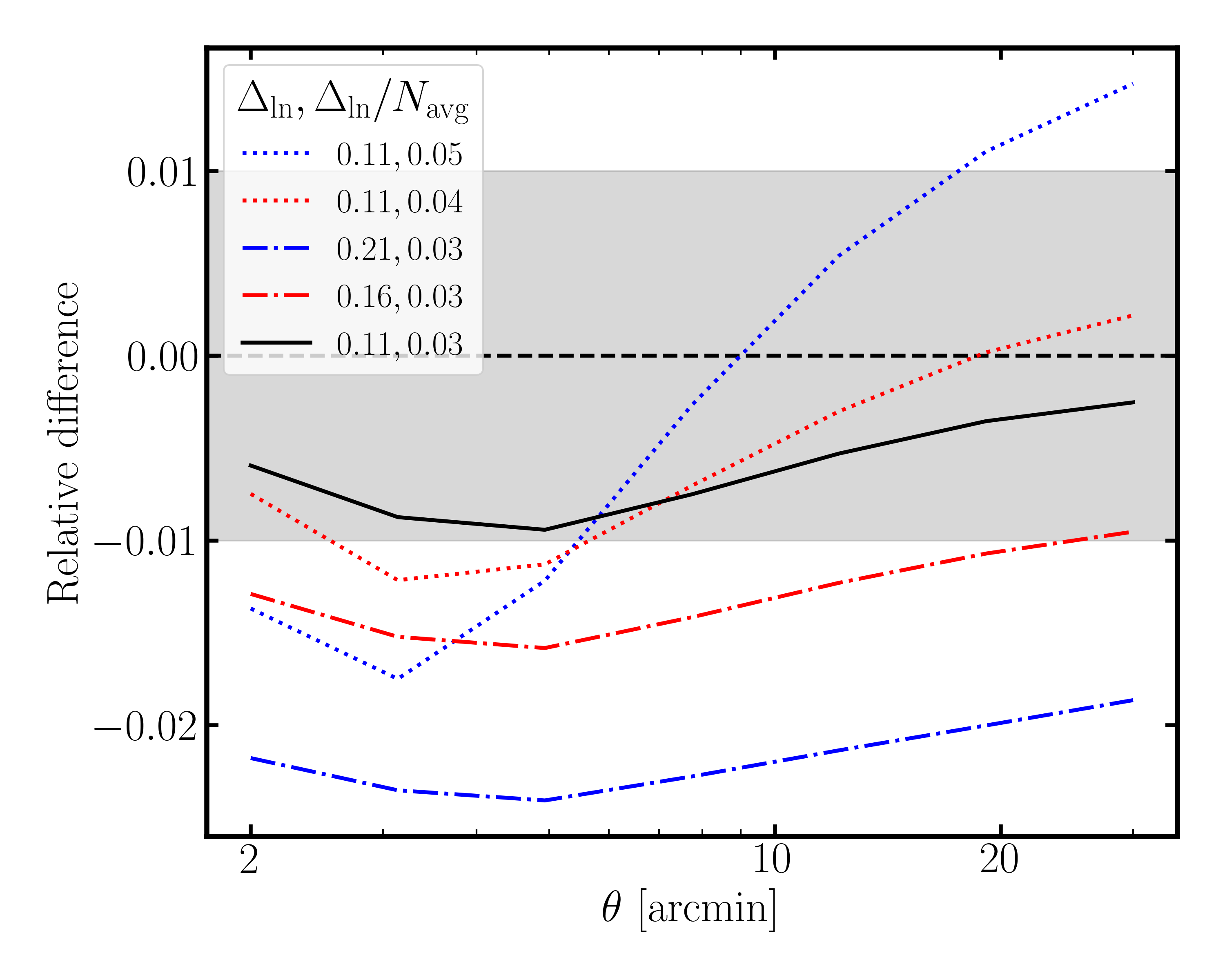}
    \caption{Relative difference of the numerical integration results for $\langle\Nap^3\Map\rangle$ performed with different values of the logarithmic bin size $\Dlog$ and $\Dlog/N_\mathrm{avg}$. The dotted lines represent calculations with fixed $\Dlog \approx 0.11$ and varying $\Dlog/N_\mathrm{avg}$, and the dash-dotted lines show calculations with varying $\Dlog$ but fixed $\Dlog/N_\mathrm{avg} \approx 0.03$. The solid black line shows the case with the smallest values for both $\Dlog$ and $\Dlog/N_\mathrm{avg}$. The numerical integrations were performed using the adaptive angular binning described in Sect. \ref{ssc:angular_binning}.}
    \label{fig:N3M_Dlog_4PCFNavg}
\end{figure}

In order to test the impact of the bin-averaged 4PCF on the numerical integration result, we perform the integration using a number of different values for $\Dlog/N_\mathrm{avg}$, while keeping the logarithmic radial bin size $\Dlog \approx 0.11$ constant. 
In Fig. \ref{fig:N3M_Dlog_4PCFNavg}, we plot the relative difference of the results with respect to the expectation from the second-order aperture statistics. 
Using the bin-averaged 4PCF heavily reduces the numerical integration error for all aperture scales and makes the error almost scale independent, as shown by the black curve with $\Dlog/N_\mathrm{avg}\approx 0.03$.
Especially at large aperture scales there is a great gain in accuracy, indicating that it is important to account for the slight differences of the radial coordinates in the galaxy configurations, which become larger with increasing aperture scale. 

\subsection{Effect of the logarithmic radial bin size}
The accuracy of the integral estimator in Eq. \eqref{eq:N3M_riemannsum} depends to a large part on the resolution in the radial coordinates that is applied in the measurement of the 4PCF. 
In Fig. \ref{fig:N3M_Dlog_4PCFNavg}, the numerical integration results are also shown for three different logarithmic radial bin sizes $\Dlog$, all performed using the same integration range and the same angular bins.
With a bin size of $\Dlog \approx 0.21$, we can already achieve results that agree with the expectation from the second-order aperture statistics within 2.5\%. 
The bin size of $\Dlog \approx 0.11$ can be considered the smallest value with which the estimation of the 4PCF is still feasible, and for this case the numerical integration results show only a sub-percent-level bias for all aperture scales. 

Although it is computationally not feasible to sample the 4PCF with an arbitrarily high resolution, we can still consider sampling the filter function with an even smaller radial bin size $\Dlog$ than presented in Fig. \ref{fig:N3M_Dlog_4PCFNavg}. 
Similarly to the procedure with the bin-averaged 4PCF in Sect. \ref{sc:bin-avg_4PCF}, we can subdivide every radial bin into a number of smaller bins.
By performing several test integrations, we find that the impact of the finer sampled filter function is rather small, adding only a $\sim 0.5\%$ additional negative bias using a radial bin size of $\Dlog \approx 0.11$ for the 4PCF. 

When we consider different aperture scales, we require a smaller logarithmic bin size in order to achieve the same level of precision as in the case of equal aperture scales.
In general, the more the ratio between the aperture scales deviates from unity, the finer the radial bins need to be to resolve the integrand appropriately. 
This is mainly due to the fact that the contributions to the aperture statistics coming from galaxy configurations with different radial coordinates are attenuated when using different aperture scales.
In this case, the main power of the integral is concentrated in configurations with similar radial coordinates which makes it more challenging to resolve the integrand appropriately with a limited bin size. 
Therefore, we cannot expect to obtain results as accurate as presented in Fig. \ref{fig:N3M_Dlog_4PCFNavg} in the case of different aperture scales.

\section{\label{sc:estimation_apstats} Estimation of aperture statistics}
\subsection{Integration of correlation functions}
In Sect. \ref{ssc:fourth-order_apstats}, we presented the mathematical framework to compute fourth-order aperture statistics by integration over the 4PCF, where one assumes a flat sky over the considered survey field. 
The fact that the correlation function can be robustly measured in the presence of survey masks makes this approach preferable to apply to real data. 
As was demonstrated in Sect. \ref{sc:measureG4L}, for Gaussian random shear and galaxy fields, the corresponding numerical integration can be performed with sufficiently high precision. 
In order to detect the connected higher-order contributions to the aperture statistics, we first need a measurement of the 4PCF.
Similar to the estimator of the shear 4PCF introduced in \cite{Porth_2025}, one can construct an efficient estimator based on a multipole decomposition of the angular dependencies of the correlation function as shown in appendix \ref{app:multipole_est}. 
However, in the following, we present an alternative, direct estimator for measuring aperture statistics that does not rely on the measurement of the 4PCF. 

\subsection{Direct estimator}
In the direct method of estimating aperture statistics, one samples the statistics on a number of apertures with centers $\{\pmb{\theta}^\mathrm{c}_i\}$ on a survey field. 
In the following, we construct an estimator for general higher-order mixed aperture moments $\langle\Nap^n \Map^m\rangle$.
The formalism applied here is similar to that used for the estimators for higher-order aperture mass moments presented in \cite{Porth_2021}.

We consider an aperture centered at position $\pmb{\theta}^\mathrm{c}$, containing the lens galaxies 
at positions $\{\pmb{\vartheta}^\mathrm{L}_i\}$ and the source galaxies at positions $\{\pmb{\vartheta}^\mathrm{S}_j\}$, and allow for different aperture scales $\thap{N}{,1},\dots,\thap{N}{,n}$ and $\thap{M}{,1},\dots,\thap{M}{,m}$, for both lens and source galaxies, respectively. 
As a direct tracer of the shear field, we denote the complex ellipticity of the $j$-th source galaxy by $\epsilon_j = \epst{,j}+\iunit\epsilon_{\times,j}$, where  $\epst{,j}$ and $\epsilon_{\times,j}$ are the tangential and cross ellipticity, respectively. 
We further allow for an ellipticity weighting with $w^\mathrm{S}_{j}$. 
The filter functions are denoted by the shorthand $U_{\thap{N}{,k};i_k} = U_{\thap{N}{,k}}(|\pmb{\vartheta}^\mathrm{L}_{i_k}-\pmb{\theta}^\mathrm{c}|)$ with $k=1,\dots,n$ for the $i_k$-th lens galaxy, and $Q_{\thap{M}{,l};j_l} = Q_{\thap{M}{,l}}(|\pmb{\vartheta}^\mathrm{S}_{j_l}-\pmb{\theta}^\mathrm{c}|)$ with $l=1,\dots,m$ for the $j_l$-th source galaxy. 
Formally, the filters do not have finite support and thus, in theory, we need to sum over all galaxies on the whole survey field to compute this estimator. 
In order to avoid spurious effects due to the finite field size, \cite{Heydenreich_2023} separated the aperture centers at least four times the maximum aperture scale from the survey boundary. 
This cut-off is applicable because the filter functions $U_\theta$ and $Q_\theta$, as given in Eq. \eqref{eq:filter_functions_definition}, drop exponentially at large separations from the aperture center, and thus contain most of their weight in a circular area $\Aap(\theta)=\pi(4\theta)^2$. 
 
The direct estimator for the mixed multiscale aperture moments of $n$-th order in $\Nap$ and $m$-th order in $\Map$, with aperture scales $\pmb{\theta}_\mathrm{ap}=(\thap{N}{,1},\dots,\thap{N}{,n},\thap{M}{,1}\dots,\thap{M}{,m})$, reads 
\be
\begin{alignedat}{1}
\label{eq:Mapm_Napn_estimator}
    \widehat{\Nap^n \Map^m}&(\pmb{\theta}^\mathrm{c};\pmb{\theta}_\mathrm{ap}) = \left[\prod_{k=1}^n\Aap(\thap{N}{,k})\right] \left[\prod_{l=1}^m\Aap(\thap{M}{,l})\right] \\
    &\times \frac{\sum_{(i_1,\dots,i_n),(j_1,\dots,j_m)} \left[\prod_{k=1}^n U_{\thap{N}{,k};i_k}\right] \cdot
    \left[\prod_{l=1}^m w^\mathrm{S}_{j_l}Q_{\thap{M}{,l};j_l}\epst{,j_l} \right] }
    { \sum_{(i_1,\dots,i_n),(j_1,\dots,j_m)} \left[\prod_{l=1}^m w^\mathrm{S}_{j_l}\right] }\;, 
\end{alignedat}
\ee
where the indices $i_1,\dots,i_n$ and $j_1,\dots,j_m$ go over all lens and source galaxies in the aperture, respectively. 
The index notation in the sums is defined as 
\be
    \sum_{(i_1,\dots,i_n)} = \sum\limits_{i_1} \sum\limits_{i_2 \neq i_1} \cdots \sum\limits_{i_1 \neq \cdots \neq i_n},
\ee
such that we avoid double countings.
Following the procedure of \cite{Porth_2021}, these sums over different indices can be further decomposed as outlined in Sect. \ref{app:directestimator}. 
Once we have obtained the estimator in Eq. \eqref{eq:Mapm_Napn_estimator} for a set of apertures on the survey field, the average over all apertures is then an estimate for the $\langle\Nap^n\Map^m\rangle$ statistics. 

\subsection{Pixelized direct estimator}
\label{ssc:FFT-method}
The direct estimator for $\langle\Nap^n\Map^m\rangle$ presented in the previous subsection requires a summation over all galaxies inside an aperture. 
When we now consider a pixelized representation of the data, we can make use of the Fast-Fourier-Transform (FFT) to perform the convolutions involved in the process of estimating aperture statistics.
The summation over the individual galaxies is numerically more exact, while the accuracy of the method based on the FFT is tied to the resolution of the underlying grid. 
However, one advantage of the pixelization is that the computational complexity of operations, like the 2-D convolution, depend only on the number of pixels and not the number of galaxies. 
While an increase in the galaxy number density results in a higher computational complexity, it leaves the speed of a discrete 2-D convolution unaffected. 
For pixelized data, this method constitutes an efficient way to compute aperture moments. 
In order to mitigate the loss of accuracy by the pixelization, it is important to note that the side length of a pixel should be substantially smaller than the minimum aperture scale considered, such that an aperture is resolved appropriately. 
The impact of the pixel size in the calculation of aperture moments is further discussed below in Sect. \ref{ssc:comparison_estimators}.

In order to be able to apply the FFT-method to compute aperture statistics, we first need to transform the estimator in Eq. \eqref{eq:Mapm_Napn_estimator} and the formalism presented in Sect. \ref{app:directestimator} from sums over indivdual galaxies to sums over pixels, as presented in Sect. \ref{app:directestimator_discrete}.

Overall, the implementation of the FFT-method to compute the general aperture statistics $\widehat{\Nap^n \Map^m}$ on a survey field can be outlined as follows: 
\begin{enumerate}
    \item Compute the filter functions $U_{\theta_\mathrm{N}}$ and $Q_{\theta_\mathrm{M}}$ on a pixel grid with the same pixel size up to a radius of at least $4\theta_\mathrm{N}$ and $4\theta_\mathrm{M}$, respectively. 
    \item Compute the quantities $\Naps{1},\dots,\Naps{n}$ and $\Maps{1},\dots,\Maps{m}$ according to Eqs. \eqref{eq:Nsm_pixel_estimator} and \eqref{eq:Msm_pixel_estimator} by convolving the galaxy number density and the shear field with the respective filter function.
    \item Multiply and sum up the $\Naps{k}$ and $\Maps{k}$ according to Eqs. \eqref{eq:Napn_Bell} and \eqref{eq:Mapm_Bell} to obtain $\widehat{\Nap^n \Map^m}$ at every pixel position serving as the aperture centers. 
    \item Average $\widehat{\Nap^n \Map^m}$ over a set of apertures to obtain $\langle\widehat{\Nap^n \Map^m}\rangle_\mathrm{r}$ for the particular realization\footnote{As different realizations of the survey field, we consider individual patches of a single full sky realization, as described in Sect. \ref{ssc:mockdata}. Given their size of about $5\times 5~\deg^2$, these can be considered to be roughly statistically independent.} of the shear and galaxy fields, denoted by the subscript r.  
\end{enumerate}

The averaging of the aperture statistics over an ensemble of apertures centered at positions $\{\pmb{\theta}^\mathrm{c}_i\}$ on the survey field, can be computed via 
\be
\langle\widehat{\Nap^n \Map^m}\rangle_\mathrm{r} 
= \frac{\sum_{\pmb{\theta}^\mathrm{c}\in \{\pmb{\theta}^\mathrm{c}_i\}} w_\mathrm{ap}(\pmb{\theta}^\mathrm{c}) \left(\widehat{\Nap^n \Map^m}\right)(\pmb{\theta}^\mathrm{c})}{\sum_{\pmb{\theta}^\mathrm{c}\in \{\pmb{\theta}^\mathrm{c}_i\}} w_\mathrm{ap}(\pmb{\theta}^\mathrm{c})}\;,
\ee
where the aperture weights $w_\mathrm{ap}$ can be chosen as presented in \cite{Porth_2021}, in order to minimize the variance of this estimator. 
When we have obtained the aperture statistics for a number of different realizations, we can further calculate the ensemble average $\langle\widehat{\Nap^n \Map^m}\rangle$ by averaging over all those realizations. 
Moreover, we can also estimate the variance $\hat{\sigma}^2\left[\langle\widehat{\Nap^n \Map^m}\rangle\right]$, and are thus able to infer the signal-to-noise ratio of the $\langle\widehat{\Nap^n \Map^m}\rangle$ statistics.

\subsection{Comparison of the estimators}
\label{ssc:comparison_estimators}
\begin{figure}
    \centering
    \includegraphics[width=\columnwidth]{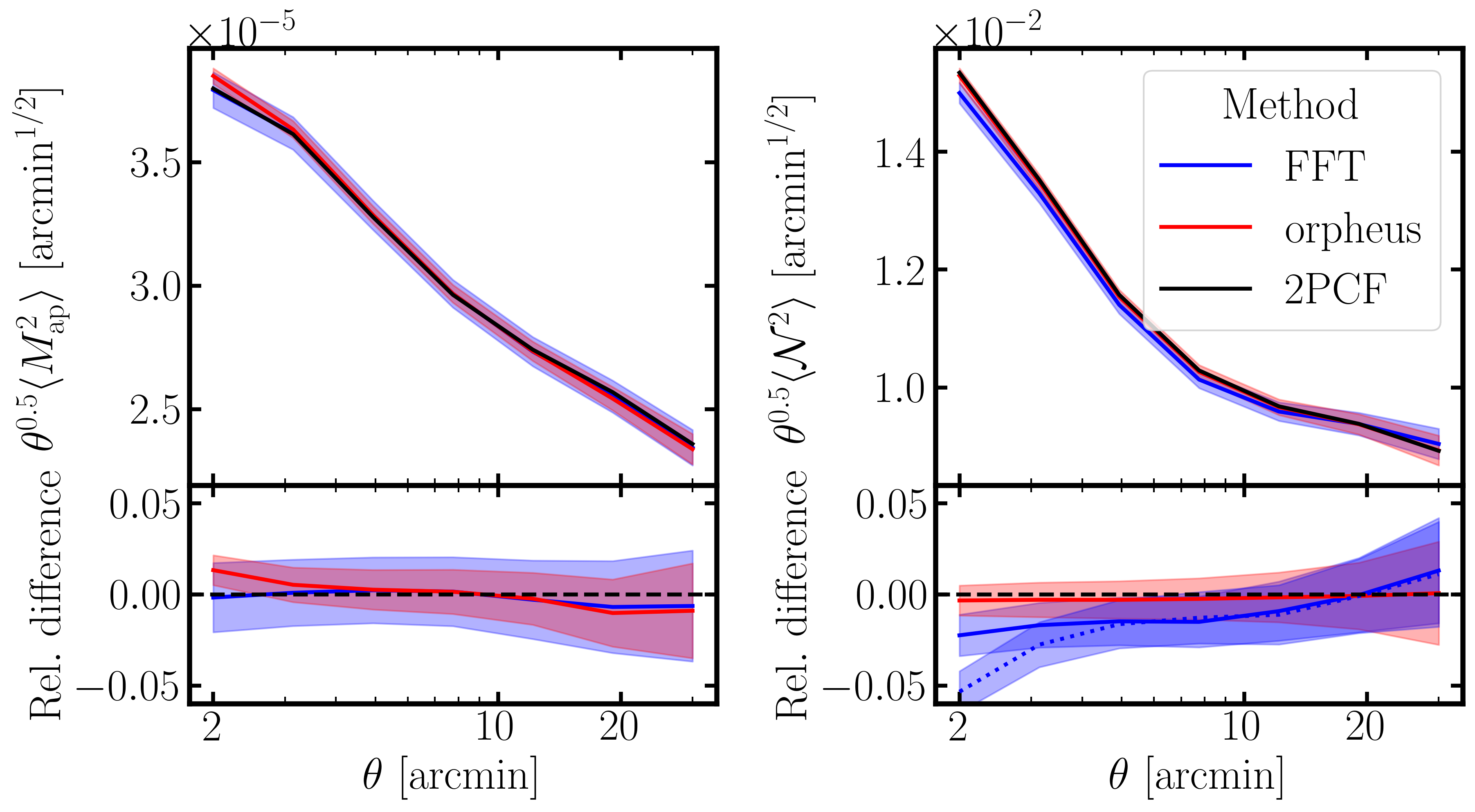}
    \caption{The aperture statistics $\langle\Map^2\rangle(\theta)$ (left) and $\langle\Nap^2\rangle(\theta)$ (right) computed with the FFT-method, as well as with \textsc{orpheus} and the direct integration of the respective 2PCFs, applied to the same mock data. The lower plots show the relative difference of the results obtained with the FFT-method and orpheus w.r.t. to the integration of the 2PCF. In the lower plot of the right panel, the solid (dotted) blue line was obtained using a pixel size of $\Delta_\mathrm{pix}=0\overset{\prime}{.}5$ ($\Delta_\mathrm{pix}=1'$) in the FFT-method. 
    The computation was performed on 1632 different mock realizations, each with an area of about $5\times5\deg^2$, and the results presented here were obtained by averaging over those. The shaded regions indicate the $1\sigma$ statistical uncertainty obtained as the standard deviation of the corresponding aperture measure over all realizations, rescaled to a sky area of $2000~\deg^2$. Lens and source galaxy populations reside in redshift bins around $z_\mathrm{lens}=0.58$ and $z_\mathrm{source}=1.67$, with mean galaxy number densities of $0.3~\mathrm{arcmin}^{-2}$ and $3~\mathrm{arcmin}^{-2}$, respectively.}
    \label{fig:Map2_Nap2_compare}
\end{figure}

In order to validate and assess the accuracy of the results obtained with the FFT-method, we apply it to the mock data described in Sect. \ref{ssc:mockdata} and compute the second-order statistics $\langle\Map^2\rangle(\theta)$ and $\langle\Nap^2\rangle(\theta)$. 
In Fig. \ref{fig:Map2_Nap2_compare}, we compare the results to the ones from the integration of the respective 2PCF, as well as the results using the \textsc{orpheus} package \footnote{\url{https://github.com/lporth93/orpheus}}. 

In the case of the second-order aperture mass statistics $\langle\Map^2\rangle(\theta)$, the results from both the FFT-method and \textsc{orpheus} are in good agreement with the integrated 2PCF on a wide range of aperture scales, showing only minor sub-per cent-level deviations. 
For $\langle\Nap^2\rangle(\theta)$, the FFT-method and \textsc{orpheus} agree well with the integrated 2PCF for aperture scales above $\sim 7'$, although the FFT-method tends to overestimate $\langle\Nap^2\rangle(\theta)$ at the largest scales towards $30'$. 
Towards small aperture scales, the FFT-method systematically underestimates $\langle\Nap^2\rangle(\theta)$ with deviations up to $\sim 2\%$ at an aperture scale of $2'$. 
On these small scales, the accuracy of the FFT-method is compromised by the finite pixel size. 
If the pixel size is not substantially smaller than the aperture scale, the filter function changes significantly over the range of a pixel, and thus approximating the filter function by evaluating it at the pixel center induces a systematic bias in the calculation of the aperture moments. 
Owing to the functional form of the filter functions, this error is expected to become larger with higher order in the aperture moments. 
We find that for the moments of $\Nap$ the evaluation of the filter function generally has a more profound impact on the result than for the $\Map$-moments. 
This bias can be mitigated by using a smaller pixel size; however, this goes hand in hand with a larger computational complexity of the FFT-convolutions involved in the process. 

During the FFT convolution, the filter functions $U_\theta$ and $Q_\theta$, as well as the lens galaxy number density field and the complex shear field, are converted to harmonic space. 
According to the Nyquist--Shannon samplimg theorem \citep{Shannon_1949}, the pixel size $\Delta_\mathrm{pix}$ sets an upper limit on the spatial frequency up to which these harmonic representations can be evaluated, which is $(2\Delta_\mathrm{pix})^{-1}$. 
Any features in the lens galaxy number density and shear field with a higher spatial frequency cannot be reconstructed accurately. 
However, the Fourier transforms of the filter functions $U_\theta$ and $Q_\theta$ are low-pass filters in harmonic space, whose attenuation of the high frequency modes of the respective field increases with the aperture scale. 
This translates to a lower limit of aperture scales down to which we can confidently compute aperture statistics, estimated by $\theta>2\Delta_\mathrm{pix}$. 
For large aperture scales the pixel size is not expected to have a significant impact on the calculation of aperture statistics, since the filter functions attenuate the signal from high frequency modes which we fail to sample correctly due to the pixelization. 

The computation of the correction terms to the aperture moments, $\Naps{k}$ and $\Maps{k}$ with $k>1$ in Eqs. \eqref{eq:Nsm_pixel_estimator} and \eqref{eq:Msm_pixel_estimator}, involves the corresponding powers of the filter functions, $U_\theta^k$ and $Q_\theta^k$. 
These are narrower peaked for $k>1$ than for $k=1$ and, in turn, they are wider in harmonic space. 
Therefore, the higher the order of the aperture moment, the stronger its correction terms are affected by the pixel size which has the largest impact on the smallest aperture scales. 

Furthermore, since the Fourier transforms of the filter functions $U_\theta$ and $Q_\theta$ are low-pass filters, their sampling is also affected by the smallest available spatial frequency. 
In real space this means that we need to sample the filter functions up to sufficiently large radii. 
For the smallest aperture scales in particular, this means that we have to evaluate the filter functions beyond the limit of $4\theta$ up to which the filter functions contain most of their weight. 
In practice, we can define a minimum radius, e.g. by setting it to $\mathrm{max}(40',4\theta)$, up to which we sample the filter functions. 

Especially for large aperture scales, the calculation of the aperture statistics is strongly affected by the selection of the apertures which we average over. 
Since the aperture moments computed in neighbouring pixels are correlated, the apertures should be selected such that they are well separated. 
Therefore, we have a smaller number of available apertures at large aperture scales, leading to a larger statistical uncertainty. 
On the other hand, sampling the aperture centers too sparsely might also give rise to aliasing effects. 



\section{\label{sc:detectability} Detection of fourth-order aperture statistics}
With the data release of stage IV weak lensing surveys, it is of particular interest to address whether we can expect a significant detection of fourth-order statistics, such as $\langle\Nap^3\Map\rangle$. 
In the following, we make use of the FFT-method introduced in the last section and apply it to simulated data sets representative of the observed data expected from stage IV surveys, in order to analyze the associated signal-to-noise ratio (S/N). 

\subsection{\label{ssc:mockdata} Mock data}
To generate more realistic mock data that results in nontrivial fourth-order correlations between the shear 
and the galaxy field, we make use of the Takahashi simulation suite \citep{Takahashi_2017}, which assumes 
a flat $\Lambda$CDM cosmology with $\Omega_\mathrm{m}=0.279$, assuming a baryon density parameter of $\Omega_\mathrm{b}=0.046$, $h=0.7$, $\sigma_8=0.82$, and $n_\mathrm{s}=0.97$. 
In particular, we obtain the matter shells and the lensing quantities for a single full-sky mock at 
healpix \citep{Gorski_2005} resolution $\texttt{nside}=8192$. 
We then use the GLASS code \citep{Tessore_2023} to create a mock galaxy distribution that  
traces the underlying matter distribution with a linear deterministic bias $b=1.2$. 
The galaxies are distributed in ten equally populated redshift bins  
with an overall number density of 
$30~\mathrm{arcmin}^{-2}$, mimicing a stage IV survey setup. 
We then retrieve the corresponding lensing quantities, i.e. position and ellipticity, such that each tracer can be used as a lens or as 
a source, where an ellipticity dispersion of $\sigma_\epsilon^2 = (0.27)^2$ is applied. 
To estimate aperture statistics on the flat sky, we cut the full sky into 1632 smaller patches with roughly equal area of $5\times 5~\si{\deg^2}$ 
and rotate the galaxies in each patch to the equator, keeping track of the change in reference direction for its ellipticity components. 

\subsection{Detection significance of $\langle\Nap^3\Map\rangle(\theta)$}
We conduct the measurement of $\langle\Nap^3\Map\rangle(\theta)$ with the FFT-method as outlined in Sect. \ref{ssc:FFT-method}, for all of the 1632 patches of the full sky, and use the individual results to estimate the ensemble average.
For the computation of the noise level, we assume the different mock patches to be statistically independent, which is, given their area of about $5\times 5~\deg^2$, justified as a rough approximation.
This allows us to obtain an estimate of the expected S/N for the corresponding aperture statistics, for which we rescale the noise level to a survey area of $2000~\deg^2$.
In the following measurements, we consider the lens galaxies in a redshift bin around $z_\mathrm{lens}=0.58$, and the source galaxies around $z_\mathrm{source}=1.67$, with mean galaxy number densities of $0.3~\mathrm{arcmin}^{-2}$ and $3~\mathrm{arcmin}^{-2}$, respectively.
While we use all available galaxies in the redshift bin of the source galaxies, we randomly mask out lens galaxies to reduce their number density.

\begin{figure}[t]
\centering
    \includegraphics[width=\columnwidth]{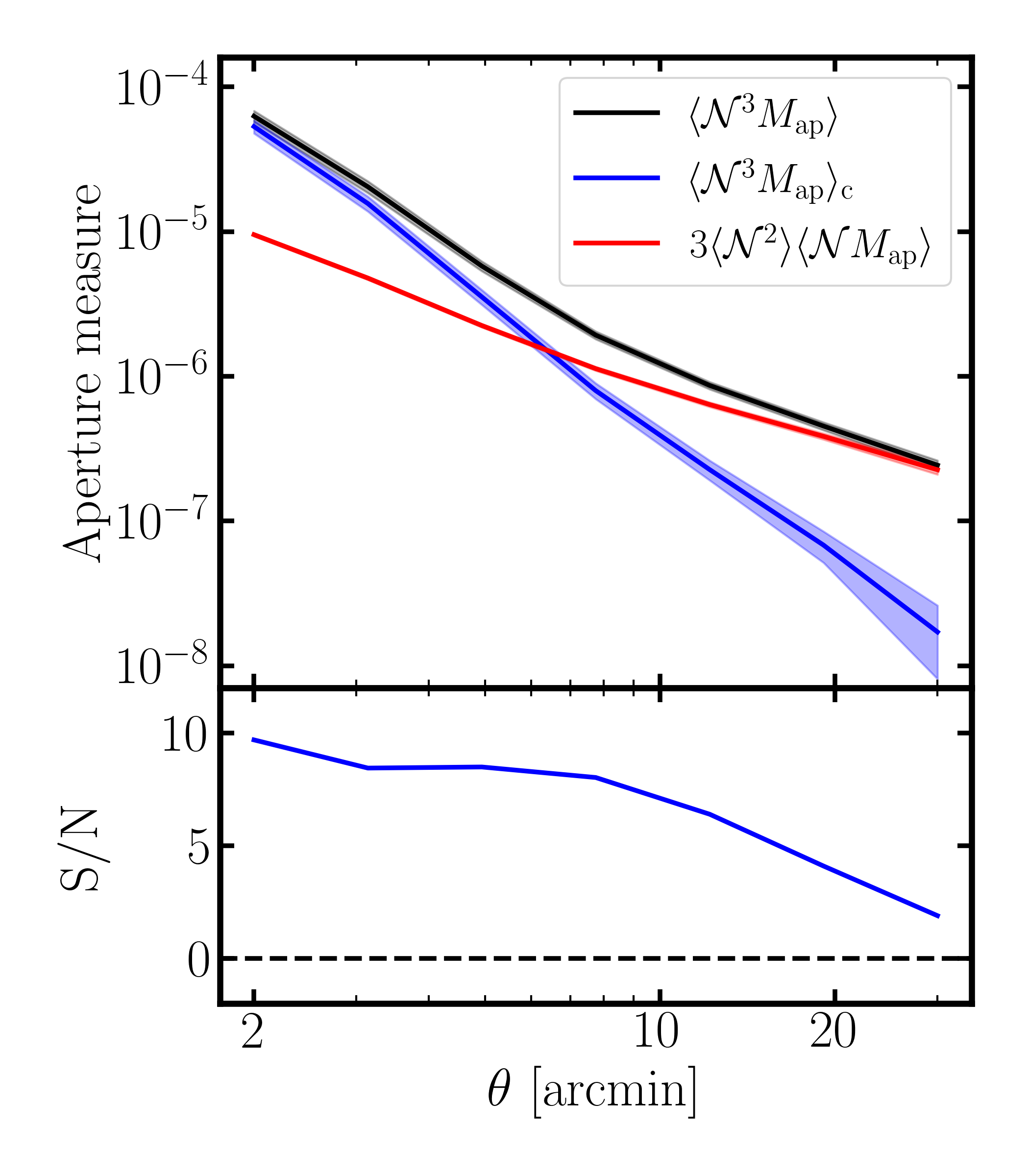}
\caption{The aperture measure $\langle\Nap^3 \Map\rangle$ as a function of the aperture scale $\theta$ calculated via the FFT-estimator on simulated mock data. In the upper plot, the black curve shows the full $\langle\Nap^3 \Map\rangle$ statistic, the blue one the connected part, and the red one the Gaussian part. The lower plot shows the S/N of the connected part at the particular aperture scales. The computation was performed on 1632 different mock realizations, each with an area of about $5\times5\deg^2$, and the results presented here were obtained by averaging over those. The shaded regions indicate the $1\sigma$ statistical uncertainty obtained as the standard deviation of the corresponding aperture measure over all realizations, rescaled to a sky area of $2000~\deg^2$. Lens and source galaxy populations reside in redshift bins around $z_\mathrm{lens}=0.58$ and $z_\mathrm{source}=1.67$, with mean galaxy number densities of $0.3~\mathrm{arcmin}^{-2}$ and $3~\mathrm{arcmin}^{-2}$, respectively. A pixel size of $\Delta_\mathrm{pix}=0\overset{\prime}{.}5$ was used in the FFT-method. 
}
\label{fig:Nap3Map_FFT_fullgaussconnected}
\end{figure}

The measurement of $\langle\Nap^3\Map\rangle(\theta)$ is shown in Fig. \ref{fig:Nap3Map_FFT_fullgaussconnected}, where we plot the full statistics, as well as its connected part and its Gaussian part, i.e. what we would expect assuming purely Gaussian random fields.
The connected part dominates the overall signal up to an aperture scale of $\theta\approx 6'$, and approaches zero towards the largest scales. 
This is where the Gaussian part dominates the full $\langle\Nap^3\Map\rangle(\theta)$ statistics which is expected since the non-Gaussian features in the matter and galaxy distribution are most prominent on small scales. 
The S/N level of the connected part at each specific aperture scale is shown in the bottom panel, being at a fairly constant level of $\mathrm{S/N}\approx 9$ at small scales and continuously declining from $\theta\approx 8'$ onwards. 
On a wide range of aperture scales, the S/N indicates that we can expect a statistically significant detection of these fourth-order aperture statistics in stage IV data.

As we have seen in Sect. \ref{ssc:comparison_estimators}, the $\Nap$-moments are underestimated by the FFT-method on small aperture scales. 
How this error propagates into the S/N of the connected part of $\langle\Nap^3\Map\rangle$ is not entirely clear, but since we expect that $\Nap^3$ is stronger negatively biased than $\Nap^2$, the S/N should be a few per cent larger than shown in Fig. \ref{fig:Nap3Map_FFT_fullgaussconnected} for small aperture scales. 


\subsection{Detection significance of $\langle\Nap^3\Map\rangle(\thap{N}{},\thap{M}{})$}
\begin{SCfigure*}[\sidecaptionrelwidth][t]
\centering
\parbox{1.45\columnwidth}{\centering
\includegraphics[width=1.45\columnwidth]{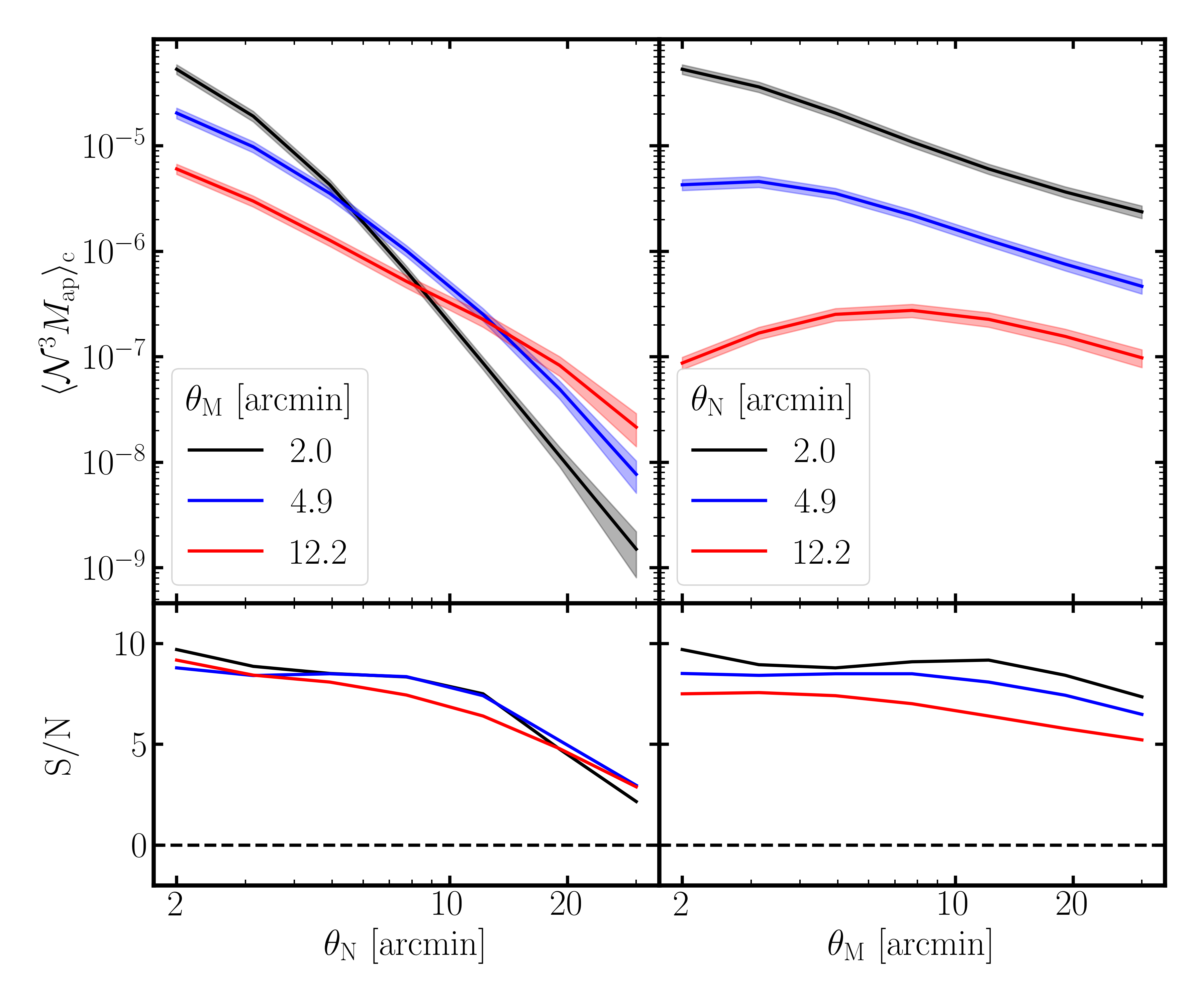}
}
\caption{The connected part $\langle\Nap^3 \Map\rangle_\mathrm{c}(\theta_\mathrm{N},\theta_\mathrm{M})$ and its detection significance as a function of the two lens and source aperture scales, $\thap{N}{}$ and $\thap{M}{}$, calculated via the FFT-estimator on simulated mock data. In the left panel, we fix the aperture scale $\theta_\mathrm{M}$ to three different values and plot $\langle\Nap^3 \Map\rangle_\mathrm{c}$ as a function of $\thap{N}{}$; in the right panel we fix $\thap{N}{}$ to the same values and plot $\langle\Nap^3 \Map\rangle_\mathrm{c}$ as a function of $\thap{M}{}$. The lower plots show the S/N level of the connected parts at the particular aperture scales. The computation was performed on 1632 different mock realizations, each with an area of $5\times5\deg^2$, and the results presented here were obtained by averaging over those. The shaded regions indicate the $1\sigma$ statistical uncertainty obtained as the standard deviation of the corresponding aperture measure over all realizations, rescaled to a sky area of $2000~\deg^2$. Lens and source galaxy populations reside in redshift bins around $z_\mathrm{lens}=0.58$ and $z_\mathrm{source}=1.67$, with mean galaxy number densities of $0.3~\mathrm{arcmin}^{-2}$ and $3~\mathrm{arcmin}^{-2}$, respectively. A pixel size of $\Delta_\mathrm{pix}=0\overset{\prime}{.}5$ was used in the FFT-method. 
}
\label{fig:Nap3Map_FFT_fullgaussconnected_multiscale}
\end{SCfigure*}

When we take into account different aperture scale configurations, additional information can be extracted from the aperture statistics. 
We apply the FFT-method to measure the $\langle\Nap^3\Map\rangle(\thap{N}{},\thap{M}{})$ statistics on the same data as in the previous section, now for equal aperture scales for the lenses, $\thap{N}{}$, but a different one for the sources, $\thap{M}{}$. 
The results are shown in Fig. \ref{fig:Nap3Map_FFT_fullgaussconnected_multiscale} where we keep one aperture scale fixed and plot $\langle\Nap^3\Map\rangle_\mathrm{c}(\thap{N}{},\thap{M}{})$ as a function of the other one.
If we fix the aperture scale for the source galaxies, $\thap{M}{}$, the S/N of the connected part as a function of $\thap{N}{}$ is very similar to what is shown in Fig. \ref{fig:Nap3Map_FFT_fullgaussconnected} for the single-aperture scale $\langle\Nap^3 \Map\rangle_\mathrm{c}(\theta)$ statistics. The S/N appears to be rather independent of the value of $\thap{M}{}$ and steadily decreases towards larger values for $\thap{N}{}$.

When we fix the aperture scale for the lens galaxies, $\thap{N}{}$, the S/N of the connected part stays at a high and almost constant level over the whole available range of $\thap{M}{}$. 
For small values of $\thap{N}{}$ in particular, the correlation between the tangential shear and close-by lens galaxy triplets belonging to the same local structure, gives major contributions to the signal of $\langle\Nap^3 \Map\rangle(\theta_\mathrm{N},\theta_\mathrm{M})$. 
In order to study the higher-order bias parameter and galaxy clustering properties on small scales, these fourth-order statistics could be a powerful discriminator in addition to its third-order counterpart $\langle\Nap^2 \Map\rangle$.

\section{Discussion}
In this paper, we have investigated the possibility of measuring fourth-order galaxy-shear correlations in the light of current stage IV surveys. 
Fourth-order statistics contain information on non-Gaussian features in the matter and galaxy distribution and will allow us to probe the relation between galaxies and the underlying mass distribution in more detail than with only second- and third-order statistics. 

Throughout this work, we focused on the correlation of the tangential shear measured around triplets of lens galaxies.
We introduced G4L statistics, naturally extending the theoretical formalism of GGL and G3L. We defined the corresponding four-point correlation function (4PCF), as well as the trispectrum, and derived the relations between both quantities. 
As a convenient statistical measure for the G4L signal, we chose to convert the 4PCF to aperture statistics, which also provide a local representation of the trispectrum.
This conversion is achieved by performing the five-dimensional integration over the 4PCF multiplied by a filter function whose analytical expression was derived. 

In Sect. \ref{sc:measureG4L}, we focused on the development of a numerical integration routine in that carries out the five-dimensional integration based on a Riemann sum. 
Assuming Gaussian shear and galaxy fields, we tested the implementation of the numerical integration and took advantage of the fact that we can express the expected results in terms of second-order statistics.
In this process, we mitigated the effect of the finite radial integration range on the accuracy of the numerical integration, and emulated the bin-averaging of the 4PCF as it would be done by a practical estimator. 
In the case of equal aperture scales, we find that the results are recovered within a one per cent-level of accuracy when we use a logarithmic radial bin size of $\Dlog \approx 0.11$, which is a feasible bin size for practical estimators of the 4PCF. 
The deviations of the integration results from the expectation is almost constant over the range of the considered aperture scales, $\theta\in[2',30']$.
The small bias induced by the numerical integration is well below the noise level that we expect for the fourth-order aperture statistics in the data of stage IV surveys such that it does not add a further substantial error to the measurement of the aperture statistics. 


Furthermore, we extended the direct estimator for aperture mass moments of \cite{Porth_2021} to estimate galaxy-mass aperture statistics.
Originally, this estimator is based on sums over discrete individual galaxies. 
In Sect. \ref{ssc:FFT-method}, we developed the pixelized counterpart of this estimator. 
By making use of FFTs, we could efficiently perform the convolutions involved in this process. 
The results for the second-order aperture statistics obtained with the FFT-method generally agree well with the results from the \textsc{orpheus} package as well as the integrated 2PCFs. 
However, for small aperture scales, the pixelization of the data induces a systematic, negative bias of a few per cent on the $\Nap$-moments, whereas this effect does not have a significant impact on the $\Map$-moments. 
Generally, a smaller pixel size reduces this bias, but also leads to higher computational costs.

By applying the FFT-method to an ensemble of mock observations constructed from the \cite{Takahashi_2017} simulation suite, we measured the $\langle\Nap^3 \Map\rangle$ statistics as the associated ensemble average over 1632 individual realizations, that cover a full sky when combined, and estimated the corresponding S/N, where we rescaled the noise level to a sky area of $2000~\deg^2$. 
The simulations mimic a typical stage IV survey setup with a mean galaxy number density of $3~\mathrm{arcmin}^{-2}$ per redshift bin.
The measurement of $\langle\Nap^3 \Map\rangle(\theta)$ shows that the connected part, which encapsulates non-Gaussian information in the matter and galaxy distribution, dominates the entire signal up to aperture scales of roughly $6'$, and is measured with a $\mathrm{S/N}\approx 9$ on small aperture scales, steadily declining from $\sim 8'$ towards larger scales.
When we discriminate between aperture scales for sources and lenses, we find that the S/N of the $\langle\Nap^3 \Map\rangle(\thap{N}{},\thap{M}{})$ statistics is mostly governed by the lens aperture scale, $\thap{N}{}$, such that we obtain a high S/N even for large values of the source galaxy aperture scale, $\thap{M}{}$.
These results indicate that we can expect to detect fourth-order aperture statistics with high significance in stage IV survey data.

Similar analyses could also be performed for the remaining galaxy-shear 4PCFs and their corresponding aperture statistics. 
Regarding $\langle\Nap^2\Map^2\rangle$ and $\langle\Nap\Map^3\rangle$, we expect the detection significance to be generally 
smaller than for $\langle\Nap^3\Map\rangle$, since there are multiple shear components involved in their respective 4PCFs that are subject to shape noise. 
The corresponding measurements can also be performed with the FFT-based estimator that can be applied to any arbitrary mixed aperture moments. 
We have shown that this estimator performs well on simulated data. 
The application to real, pixelized data will most likely induce systematic biases on any measurement due to masks on the survey field that are not present in the simulated data used here.
The impact of these biases would need to be investigated by comparing the measurements of the FFT-method to the results obtained via the direct integration of the corresponding correlation function and \textsc{orpheus}, which will suffer in a similar way due to masks. 

The fact that fourth-order galaxy-shear statistics will be detectable in stage IV survey data shows that they can constitute an extension to analyses on third-order statistics in order to probe the non-Gaussian features in the Universe's matter and galaxy distribution. 
Whether the consideration of fourth-order galaxy-shear statistics will add new information and may break degeneracies when added to the combined analyses of second- and third-order statistics, has to be shown in future work. 
If that is the case, the high-precision observations of stage IV surveys make it possible to investigate galaxy biasing and constrain models of galaxy evolution in more detail by taking fourth-order statistics into account.
Even if completely degenerate with second- and third-order statistics, fourth-order statistics could still be used to study nuisance parameters of such models.

%
%

\begin{acknowledgements}
LP acknowledges support from the DLR grant 50QE2302.
\end{acknowledgements}

%
%

\bibliography{Euclid, Q1} 

%

\begin{appendix}
  \onecolumn 
  
\section{\label{app:A} Further fourth-order statistics }


\subsection{\label{app:A_4PCFs} Four-point correlation functions}
In addition to the galaxy-galaxy-galaxy-shear 4PCF introduced in Eq. \eqref{eq:Ggggk_definition}, the remaining 4PCFs are provided here.

\begin{figure}[t]
    \label{fig:gsss_ggss_sketch}
    \centering
    \usetikzlibrary {positioning}
\usetikzlibrary{arrows}
\begin{tikzpicture}
    \pgfmathsetmacro{\xone}{cos(\firstdir)}%
    \pgfmathsetmacro{\yone}{sin(\firstdir)}%
    \pgfmathsetmacro{\xtwo}{cos(\seconddir)}%
    \pgfmathsetmacro{\ytwo}{sin(\seconddir)}%
    \pgfmathsetmacro{\xthree}{cos(\thirddir)}%
    \pgfmathsetmacro{\ythree}{sin(\thirddir)}%
    \coordinate (a) at ($\length*(\xone,\yone)$);
    \coordinate (b) at ($\length*(\xtwo,\ytwo)$);
    \coordinate (c) at ($\length*(\xthree,\ythree)$);
    \coordinate (o) at (0,0);
    
    \filldraw[black] (0,0) circle (2pt) node [below,pos=0] {\Large$\pmb{\theta}_4$};
    
    \draw[line width=1pt] [->] (0,0) -- (a) node [below,pos=0.7] {\Large$\pmb{\vartheta}_1$};
    \draw[line width=1pt] [->] (0,0) -- (b) node [right,pos=0.8] {\Large$\pmb{\vartheta}_2$};
    \draw[line width=1pt] [->] (0,0) -- (c) node [left,pos=0.5] {\Large$\pmb{\vartheta}_3$};

    \draw[line width=2pt] ($\length*(\xone,\yone) - 0.3*(-\yone,\xone)$) -- ($\length*(\xone,\yone) + 0.3*(-\yone,\xone)$) node [below,pos=0] {\Large$\pmb{\theta}_1$};
    \draw[line width=2pt] ($\length*(\xtwo,\ytwo) - 0.3*(-\ytwo,\xtwo)$) -- ($\length*(\xtwo,\ytwo) + 0.3*(-\ytwo,\xtwo)$) node [left,pos=1] {\Large$\pmb{\theta}_2$};
    \filldraw[black] (c) circle (2pt) node (x) {};
    \node[below left=10pt, label=\Large$\pmb{\theta}_3$] at (c) {};
    
    \draw [->] ($2*(\xone,\yone)$)  arc (\firstdir:\seconddir:2) node [above,pos=0.5] {\Large$\phi_{12}$};
    \draw [->] ($2*(\xtwo,\ytwo)$)  arc (\seconddir:\thirddir:2) node [above,pos=0.5] {\Large$\phi_{23}$};
    \draw [->] ($3*(\xone,\yone)$)  arc (\firstdir:\thirddir:3) node [above left,pos=0.5] {\Large$\phi_{13}$};

    \draw [dashed] (a) -- ($1.2*\length*(\xone,\yone)$);
    \draw [dashed] (a) -- ($\length*(\xone,\yone)+(0.2*\length,0)$);
    \draw [dashed] (b) -- ($1.2*\length*(\xtwo,\ytwo)$);
    \draw [dashed] (b) -- ($\length*(\xtwo,\ytwo)+(0.2*\length,0)$);
    \draw [dashed] (c) -- ($1.2*\length*(\xthree,\ythree)$);
    \draw [dashed] (c) -- ($\length*(\xthree,\ythree)+(0.2*\length,0)$);

    \draw [->] ($\length*(\xone,\yone)+(0.1*\length,0)$)  arc (0:\firstdir:0.45) node [right,pos=0.5] {\Large$\varphi_1$};
    \draw [->] ($\length*(\xtwo,\ytwo)+(0.1*\length,0)$)  arc (0:\seconddir:0.45) node [above right,pos=0.5] {\Large$\varphi_2$};
    \draw [->] ($\length*(\xthree,\ythree)+(0.1*\length,0)$)  arc (0:\thirddir:0.45) node [above right,pos=0.5] {\Large$\varphi_3$};

\end{tikzpicture}
    \hfill
    \usetikzlibrary {positioning}
\usetikzlibrary{arrows}
\begin{tikzpicture}
    \pgfmathsetmacro{\xone}{cos(\firstdir)}%
    \pgfmathsetmacro{\yone}{sin(\firstdir)}%
    \pgfmathsetmacro{\xtwo}{cos(\seconddir)}%
    \pgfmathsetmacro{\ytwo}{sin(\seconddir)}%
    \pgfmathsetmacro{\xthree}{cos(\thirddir)}%
    \pgfmathsetmacro{\ythree}{sin(\thirddir)}%
    \coordinate (a) at ($\length*(\xone,\yone)$);
    \coordinate (b) at ($\length*(\xtwo,\ytwo)$);
    \coordinate (c) at ($\length*(\xthree,\ythree)$);
    \coordinate (o) at (0,0);
    
    \filldraw[black] (0,0) circle (2pt) node [below,pos=0] {\Large$\pmb{\theta}_4$};
    
    \draw[line width=1pt] [->] (0,0) -- (a) node [below,pos=0.7] {\Large$\pmb{\vartheta}_1$};
    \draw[line width=1pt] [->] (0,0) -- (b) node [right,pos=0.8] {\Large$\pmb{\vartheta}_2$};
    \draw[line width=1pt] [->] (0,0) -- (c) node [left,pos=0.5] {\Large$\pmb{\vartheta}_3$};

    \draw[line width=2pt] ($\length*(\xone,\yone) - 0.3*(-\yone,\xone)$) -- ($\length*(\xone,\yone) + 0.3*(-\yone,\xone)$) node [below,pos=0] {\Large$\pmb{\theta}_1$};
    \draw[line width=2pt] ($\length*(\xtwo,\ytwo) - 0.3*(-\ytwo,\xtwo)$) -- ($\length*(\xtwo,\ytwo) + 0.3*(-\ytwo,\xtwo)$) node [left,pos=1] {\Large$\pmb{\theta}_2$};
    \draw[line width=2pt] ($\length*(\xthree,\ythree) - 0.3*(-\ythree,\xthree)$) -- ($\length*(\xthree,\ythree) + 0.3*(-\ythree,\xthree)$) node [left,pos=0.7] {\Large$\pmb{\theta}_3$};
    
    \draw [->] ($2*(\xone,\yone)$)  arc (\firstdir:\seconddir:2) node [above,pos=0.5] {\Large$\phi_{12}$};
    \draw [->] ($2*(\xtwo,\ytwo)$)  arc (\seconddir:\thirddir:2) node [above,pos=0.5] {\Large$\phi_{23}$};
    \draw [->] ($3*(\xone,\yone)$)  arc (\firstdir:\thirddir:3) node [above left,pos=0.5] {\Large$\phi_{13}$};

    \draw [dashed] (a) -- ($1.2*\length*(\xone,\yone)$);
    \draw [dashed] (a) -- ($\length*(\xone,\yone)+(0.2*\length,0)$);
    \draw [dashed] (b) -- ($1.2*\length*(\xtwo,\ytwo)$);
    \draw [dashed] (b) -- ($\length*(\xtwo,\ytwo)+(0.2*\length,0)$);
    \draw [dashed] (c) -- ($1.2*\length*(\xthree,\ythree)$);
    \draw [dashed] (c) -- ($\length*(\xthree,\ythree)+(0.2*\length,0)$);

    \draw [->] ($\length*(\xone,\yone)+(0.1*\length,0)$)  arc (0:\firstdir:0.45) node [right,pos=0.5] {\Large$\varphi_1$};
    \draw [->] ($\length*(\xtwo,\ytwo)+(0.1*\length,0)$)  arc (0:\seconddir:0.45) node [above right,pos=0.5] {\Large$\varphi_2$};
    \draw [->] ($\length*(\xthree,\ythree)+(0.1*\length,0)$)  arc (0:\thirddir:0.45) node [above right,pos=0.5] {\Large$\varphi_3$};

\end{tikzpicture}
    \caption{A galaxy-galaxy-shear-shear configuration (left panel) which $\ggnnCF^\pm$ is sensitive to and a galaxy-shear-shear-shear configuration (right panel) which $\gggnCF^\pm$ is sensitive to.}
\end{figure}
First, we examine the configuration of the shear measured at the position of two background galaxies correlated with the position of a pair of foreground galaxies, as depicted in the left panel of Fig. \ref{fig:gsss_ggss_sketch}.
Initially, we project the rotated shear components in the directions $\varphi_i$ of the separation vectors $\pmb{\vartheta}_i$ to the position of the foreground galaxy at $\pmb{\theta}_4$. 
In analogy to the shear 2PCFs, $\xi_\pm$, 
we can define the two 4PCFs, $\ggnnCF^\pm$ by
\be
\begin{alignedat}{1}
    \ggnnCF^+(\pmb{\vartheta}_1,\pmb{\vartheta}_2,\pmb{\vartheta}_3) &= \ggnnCF^+(\vartheta_1,\vartheta_2,\vartheta_3,\phi_{12},\phi_{23}) = \langle\gamma(\pmb{\theta}_1;\varphi_1)~\gamma^*(\pmb{\theta}_2;\varphi_2)~\kappa_\mathrm{g}(\pmb{\theta}_3)~\kappa_\mathrm{g}(\pmb{\theta}_4)\rangle\;, \\
    \ggnnCF^-(\pmb{\vartheta}_1,\pmb{\vartheta}_2,\pmb{\vartheta}_3) &= \ggnnCF^-(\vartheta_1,\vartheta_2,\vartheta_3,\phi_{12},\phi_{23}) = \langle\gamma(\pmb{\theta}_1;\varphi_1)~\gamma(\pmb{\theta}_2;\varphi_2)~\kappa_\mathrm{g}(\pmb{\theta}_3)~\kappa_\mathrm{g}(\pmb{\theta}_4)\rangle\;,
\end{alignedat}
\ee
where $\pmb{\vartheta}_i = \pmb{\theta}_i-\pmb{\theta}_4$ denotes the angular separation vector from one background galaxy to the foreground galaxy at position $\pmb{\theta}_4$ with polar angles $\varphi_i$. 
Two of those vectors, $\pmb{\vartheta}_i$ and $\pmb{\vartheta}_j$, enclose an angle $\phi_{ij} = \varphi_j - \varphi_i$ which follow the simple relation $\phi_{13}=\phi_{12}+\phi_{23}$.
As before, we define the modified correlators as
\be
\begin{alignedat}{1}
    \tilde{G}^+_{\gamma\gamma\mathrm{gg}}(\vartheta_1,\vartheta_2,\vartheta_3,\phi_{12},\phi_{23}) &= \frac{1}{\Bar{N}^2} \langle\gamma(\pmb{\theta}_1;\varphi_1)~\gamma^*(\pmb{\theta}_2;\varphi_2)~N(\pmb{\theta}_3)~N(\pmb{\theta}_4)\rangle\;,\\
    \tilde{G}^-_{\gamma\gamma\mathrm{gg}}(\vartheta_1,\vartheta_2,\vartheta_3,\phi_{12},\phi_{23}) &= \frac{1}{\Bar{N}^2} \langle\gamma(\pmb{\theta}_1;\varphi_1)~\gamma(\pmb{\theta}_2;\varphi_2)~N(\pmb{\theta}_3)~N(\pmb{\theta}_4)\rangle\;,
\end{alignedat}
\ee
which we can rewrite in terms of $\ggnnCF^\pm$ and lower-order correlators. 
In the following, we use the fact that the rotated shear with projection angle $\phi$ can be expressed through the rotated shear 
with a different projection angle $\phi'$ according to $\gamma(\pmb{\theta};\phi) = \gamma(\pmb{\theta};\phi')~\eexp^{2\iunit(\phi'-\phi)}$. 
Furthermore, we denote the polar angle of the separation vector $\pmb{\Delta\theta}_{ij}=\pmb{\vartheta}_j- \pmb{\vartheta}_i$ as $\psi_{ij}$. 
With these relations, we can write
\be
\label{eq:ggnn_modified}
\begin{alignedat}{1}
    \tilde{G}^+_{\gamma\gamma\mathrm{gg}}(\vartheta_1,\vartheta_2,\vartheta_3,\phi_{12},\phi_{23})
    &= \ggnnCF^+(\vartheta_1,\vartheta_2,\vartheta_3,\phi_{12},\phi_{23}) + \xi_+(\Delta\theta_{12})~\eexp^{2\iunit\phi_{12}} + G^+_{\gamma\gamma\mathrm{g}}(\vartheta_1,\vartheta_2, \phi_{12}) \\
    &\quad + G^+_{\gamma\gamma\mathrm{g}}(\Delta\theta_{13},\Delta\theta_{23},\psi_{23}-\psi_{13})~\frac{\left(\vartheta_3~\eexp^{\iunit(\phi_{12}+\phi_{23})} - \vartheta_1\right)^2}{\Delta\theta_{13}^2}~\frac{\left(\vartheta_3~\eexp^{-\iunit{\phi_{23}}} - \vartheta_2\right)^2}{\Delta\theta_{23}^2}\;,
\end{alignedat}
\ee
where we note that the polar angle $\psi_{ij}$ follows the relation 
$\eexp^{\iunit\psi_{ij}}=\left(\vartheta_j~\eexp^{\iunit\varphi_j} - \vartheta_i~\eexp^{\iunit\varphi_i}\right)/\Delta\theta_{ij}$, with $\Delta\theta_{ij}=\sqrt{\vartheta_i^2+\vartheta_j^2-2\vartheta_i\vartheta_j\cos\phi_{ij}}$. 
In the definition of the 3PCF, $G^+_{\gamma\gamma\mathrm{g}} = \langle\gamma~\gamma^*~\kappa_\mathrm{g}\rangle$, in \cite{Schneider_2005}, the shear components are projected towards the foreground galaxy. 
This is the origin of the phase factor for the second 3PCF in Eq. \eqref{eq:ggnn_modified}.
Analogously, the shears have to be projected towards each other's position in order to apply the definition of the shear 2PCF, $\xi_+=\langle\gamma~\gamma^*\rangle$. 
Changing the initial projection of the shear components in the definition of $\ggnnCF^+$ will also affect the phase factors, in general. 

In a similar manner, it follows that 
\be
\begin{alignedat}{1}
    \tilde{G}^-_{\gamma\gamma\mathrm{gg}}(\vartheta_1,\vartheta_2,\vartheta_3,\phi_{12},\phi_{23}) 
    &= \ggnnCF^-(\vartheta_1,\vartheta_2,\vartheta_3,\phi_{12},\phi_{23}) + G^-_{\gamma\gamma\mathrm{g}}(\vartheta_1,\vartheta_2, \phi_{12}) + \xi_-(\Delta\theta_{12})~\frac{\left(\vartheta_2~\eexp^{\iunit\phi_{12}/2} - \vartheta_1~\eexp^{-\iunit\phi_{12}/2}\right)^4}{\Delta\theta_{12}^4} \\
    &\quad + G^-_{\gamma\gamma\mathrm{g}}(\Delta\theta_{13},\Delta\theta_{23},\psi_{23}-\psi_{13})~\frac{\left(\vartheta_3~\eexp^{\iunit(\phi_{12}+\phi_{23})} - \vartheta_1\right)^2}{\Delta\theta_{13}^2}~\frac{\left(\vartheta_3~\eexp^{\iunit{\phi_{23}}} - \vartheta_2\right)^2}{\Delta\theta_{23}^2}\;.
\end{alignedat}
\ee
These results show that the modified galaxy-galaxy-shear-shear correlators are given by the correlators 
$\ggnnCF^\pm$ with additional contributions from second- and third-order correlators multiplied 
by their respective phase factors originating in the shear projection. 

We also consider the case of three shears correlated with the position of a single background galaxy. 
Here we need to adapt to the definition of the natural components, $\Gamma_i$, \citep[cf.][]{Schneider_2003} of the shear 3PCFs 
and define the four 4PCFs as
\be
\label{eq:Gkkkg_naturalcomp_def}
\begin{alignedat}{2}
    \gggnCF^{(0)}(\pmb{\vartheta}_1,\pmb{\vartheta}_2,\pmb{\vartheta}_3) 
    &= \langle\gamma(\pmb{\theta}_1;\varphi_1)~\gamma(\pmb{\theta}_2;\varphi_2)~\gamma(\pmb{\theta}_3;\varphi_3)~\kappa_\mathrm{g}(\pmb{\theta}_4)\rangle\;, \qquad
    &\gggnCF^{(1)}(\pmb{\vartheta}_1,\pmb{\vartheta}_2,\pmb{\vartheta}_3) 
    = \langle\gamma^*(\pmb{\theta}_1;\varphi_1)~\gamma(\pmb{\theta}_2;\varphi_2)~\gamma(\pmb{\theta}_3;\varphi_3)~\kappa_\mathrm{g}(\pmb{\theta}_4)\rangle\;, \\[0.4cm]
    \gggnCF^{(2)}(\pmb{\vartheta}_1,\pmb{\vartheta}_2,\pmb{\vartheta}_3) 
    &= \langle\gamma(\pmb{\theta}_1;\varphi_1)~\gamma^*(\pmb{\theta}_2;\varphi_2)~\gamma(\pmb{\theta}_3;\varphi_3)~\kappa_\mathrm{g}(\pmb{\theta}_4)\rangle\;, \qquad
    &\gggnCF^{(3)}(\pmb{\vartheta}_1,\pmb{\vartheta}_2,\pmb{\vartheta}_3) 
    = \langle\gamma(\pmb{\theta}_1;\varphi_1)~\gamma(\pmb{\theta}_2;\varphi_2)~\gamma^*(\pmb{\theta}_3;\varphi_3)~\kappa_\mathrm{g}(\pmb{\theta}_4)\rangle\;,
\end{alignedat}
\ee
where we initially project the shears in the direction $\varphi_i$ towards the foreground galaxy at $\pmb{\theta}_4$. 

In analogy to Eq. \eqref{eq:modifiedGgggk_definition}, the modified correlators are defined by replacing $\kappa_\mathrm{g}$ with $N/\Bar{N}$ in Eq. \eqref{eq:Gkkkg_naturalcomp_def} and contain a contribution from the shear 3PCFs.
When we apply $\times$-projection \citep[cf.][]{Porth_2024} to the natural components, we choose the shear at position $\pmb{\theta}_1$ as the new ``origin'' and project the other two shears in the direction of their separation vector $\pmb{\Delta\theta}_{1i}=\pmb{\vartheta}_i-\pmb{\vartheta}_1$ with polar angle $\psi_{1i}$.
The shear at $\pmb{\theta}_1$ is projected along the bisection of $\pmb{\Delta\theta}_{12}$ and $\pmb{\Delta\theta}_{13}$.
For the 0-component we have 
\be
\begin{alignedat}{1}
\tilde{G}^{(0)}_{\gamma\gamma\gamma\mathrm{g}} = \gggnCF^{(0)} - \langle\gamma(\pmb{\theta}_1)~\gamma(\pmb{\theta}_2)~\gamma(\pmb{\theta}_3)\rangle~\eexp^{-2\iunit (\varphi_1+\varphi_2+\varphi_3)}
&= \gggnCF^{(0)} + \bigg\langle\gamma\left(\pmb{\theta}_1;\frac{\psi_{12}+\psi_{13}}{2}\right)~\gamma(\pmb{\theta}_2;\psi_{12})~ \gamma(\pmb{\theta}_3;\psi_{13})\bigg\rangle~\eexp^{3\iunit(\psi_{12}+\psi_{13})}~\eexp^{-2\iunit (\varphi_1+\varphi_2+\varphi_3)} \\
&= \gggnCF^{(0)} + \Gamma^\times_0~\frac{\left(\vartheta_2~\eexp^{\iunit\phi_{12}}-\vartheta_1\right)^3}{\Delta\theta_{12}^3}~\frac{\left(\vartheta_3~\eexp^{\iunit(\phi_{12}+\phi_{23})}-\vartheta_1\right)^3}{\Delta\theta_{13}^3}~\eexp^{-2\iunit (2\phi_{12}+\phi_{23})}\;.
\end{alignedat}
\ee
Similarly, we can write the remaining correlation functions as 
\be
\begin{alignedat}{2}
    \tilde{G}^{(1)}_{\gamma\gamma\gamma\mathrm{g}}(\pmb{\vartheta}_1,\pmb{\vartheta}_2,\pmb{\vartheta}_3) 
    &= \frac{1}{\Bar{N}}\langle\gamma^*(\pmb{\theta}_1;\varphi_1)~\gamma(\pmb{\theta}_2;\varphi_2)~\gamma(\pmb{\theta}_3;\varphi_3)~N(\pmb{\theta}_4)\rangle = \gggnCF^{(1)} + \Gamma^\times_1~\frac{\vartheta_2~\eexp^{\iunit\phi_{12}}-\vartheta_1}{\Delta\theta_{12}}~\frac{\vartheta_3~\eexp^{\iunit(\phi_{12}+\phi_{23})}-\vartheta_1}{\Delta\theta_{13}}~\eexp^{-2\iunit (2\phi_{12}+\phi_{23})}\;, \\
    \tilde{G}^{(2)}_{\gamma\gamma\gamma\mathrm{g}}(\pmb{\vartheta}_1,\pmb{\vartheta}_2,\pmb{\vartheta}_3) 
    &= \frac{1}{\Bar{N}}\langle\gamma(\pmb{\theta}_1;\varphi_1)~\gamma^*(\pmb{\theta}_2;\varphi_2)~\gamma(\pmb{\theta}_3;\varphi_3)~N(\pmb{\theta}_4)\rangle = \gggnCF^{(2)} + \Gamma^\times_2~\frac{\vartheta_2~\eexp^{-\iunit\phi_{12}}-\vartheta_1}{\Delta\theta_{12}}~\frac{\left(\vartheta_3~\eexp^{\iunit(\phi_{12}+\phi_{23})}-\vartheta_1\right)^3}{\Delta\theta_{13}^3}~\eexp^{-2\iunit\phi_{23}}\;, \\
    \tilde{G}^{(3)}_{\gamma\gamma\gamma\mathrm{g}}(\pmb{\vartheta}_1,\pmb{\vartheta}_2,\pmb{\vartheta}_3) 
    &= \frac{1}{\Bar{N}}\langle\gamma(\pmb{\theta}_1;\varphi_1)~\gamma(\pmb{\theta}_2;\varphi_2)~\gamma^*(\pmb{\theta}_3;\varphi_3)~N(\pmb{\theta}_4)\rangle = \gggnCF^{(3)} + \Gamma^\times_3~\frac{\left(\vartheta_2~\eexp^{\iunit\phi_{12}}-\vartheta_1\right)^3}{\Delta\theta_{12}^3}~\frac{\vartheta_3~\eexp^{-\iunit(\phi_{12}+\phi_{23})}-\vartheta_1}{\Delta\theta_{13}}~\eexp^{2\iunit\phi_{23}}\;.
\end{alignedat}
\ee


\subsection{\label{app:A_4PCF_trispectra} Relations between the 4PCF and the trispectrum}
The 4PCFs and the trispectra are defined through the Fourier transforms of the convergence and galaxy number density contrast, $\hat{\kappa}$ and $\hat{\kappa}_\mathrm{g}$. 
By using the relation $\hat{\gamma}_\mathrm{c}(\pmb{\ell})=\hat{\kappa}(\pmb{\ell})~\eexp^{2\iunit\beta}$, where $\beta$ is the polar angle of $\pmb{\ell}$, between shear and convergence in Fourier space, one can express the 4PCFs in terms of the trispectra and vice versa. 
The calculations below are done in a similar fashion to the derivation of the relations between the 3PCFs and the bispectra presented in \cite{Schneider_2005}.

In the case of the galaxy-galaxy-galaxy-shear correlation, this relation includes the general projection angle $\zeta$ of the only shear component of the galaxy quadruplet, such that we can write
\be
    \nnngCF(\pmb{\vartheta}_1,\pmb{\vartheta}_2,\pmb{\vartheta}_3) = - \left(\prod\limits_{i=1}^3 \int_0^\infty \frac{\dd^2 \ell_i}{(2\pi)^2}\right) 
    \eexp^{-\iunit(\pmb{\ell}_1\cdot\pmb{\vartheta}_1 + \pmb{\ell}_2\cdot\pmb{\vartheta}_2 + \pmb{\ell}_3\cdot\pmb{\vartheta}_3)}~ 
    \eexp^{2\iunit(\beta_4-\zeta)}~t_{\mathrm{ggg}\kappa}(\pmb\ell_1,\pmb\ell_2,\pmb\ell_3;-\pmb\ell_1-\pmb\ell_2-\pmb\ell_3)\;,
\ee
where, due to the delta function in the definition of the trispectrum, $\eexp^{\iunit\beta_4}$ follows 
\begin{equation}
    \ell_4~\eexp^{\iunit\beta_4} = -\sum_{k=1}^3 \ell_k~\eexp^{\iunit\beta_k}, \quad\text{with}\quad
    \ell_4^2 = \ell_1^2 + \ell_2^2 + \ell_3^2 + 2\ell_1\ell_2\cos(\psi_{12}) + 2\ell_1\ell_3\cos(\psi_{12}+\psi_{23}) + 2\ell_2\ell_3\cos(\psi_{23})\;,
\end{equation}
where $\psi_{ij} = \beta_j-\beta_i$.
This integral can be further simplified by noting that, similar to the 4PCFs, the trispectra only depend on the lengths of three $\pmb{\ell}_i$ vectors and two angles $\psi_{ij}$ enclosed by those. 
Thus we can choose a transformation by defining $\eta=(\beta_1 + \beta_3)/2$, analogously to the choice of $\zeta=(\varphi_1 + \varphi_3)/2$. 
This allows us to write the angles $\beta_i$ in terms of $\eta,~\psi_{12}$ and $\psi_{23}$ as
\be
    \beta_1 = \eta - \psi_{12}/2 - \psi_{23}/2, \quad 
    \beta_2 = \eta + \psi_{12}/2 - \psi_{23}/2, \text{ and }
    \beta_3 = \eta + \psi_{12}/2 + \psi_{23}/2\;.
\ee
Furthermore, by writing the product $\pmb{\ell}_1\cdot\pmb{\vartheta}_1$ using the complex representations of the 2D-vectors, we can rewrite the expression in the exponential of the Fourier transformation as 
\be
\pmb{\ell}_1\cdot\pmb{\vartheta}_1 + \pmb{\ell}_2\cdot\pmb{\vartheta}_2 + \pmb{\ell}_3\cdot\pmb{\vartheta}_3 
= \left[A\cos(\eta-\zeta-\nu)\right](\ell_1\vartheta_1,\ell_2\vartheta_2,\ell_3\vartheta_3,\phi_{12}-\psi_{12},\phi_{23}-\psi_{23})\;,
\ee
where we define the dimensionless quantities 
\be
\begin{alignedat}{1}
    A(x_1,x_2,x_3,\alpha_{12},\alpha_{23}) &= x_1^2 + x_2^2 + x_3^2 + 2x_1x_2\cos(\alpha_{12}) + 2x_1x_3\cos(\alpha_{12}+\alpha_{23}) + 2x_2x_3\cos(\alpha_{23}), \\
    \eexp^{\iunit\nu(x_1,x_2,x_3,\alpha_{12},\alpha_{23})} &= \frac{1}{A(x_1,x_2,x_3,\alpha_{12},\alpha_{23})}\Bigg[x_1\exp\left(-\frac{\iunit}{2}(\alpha_{12} + \alpha_{23})\right) + x_2\exp\left(\frac{\iunit}{2}(\alpha_{12} - \alpha_{23})\right)
    + x_3\exp\left(\frac{\iunit}{2}(\alpha_{12} + \alpha_{23})\right)\Bigg]\;.
\end{alignedat}
\ee
With the above transformations and by using the $n$-th order Bessel function of the first kind $J_n(x)$, 
the relations between the 4PCF and the trispectrum read
\be
\begin{alignedat}{3}
    &\nnngCF(\vartheta_1,\vartheta_2,\vartheta_3,\phi_{12},\phi_{23}) 
    &&= \left(\prod\limits_{i=1}^3 \int_0^\infty \frac{\dd\ell_i~\ell_i}{2\pi}\right) \int_0^{2\pi}\frac{\dd\psi_{12}}{2\pi}\int_0^{2\pi}\frac{\dd\psi_{23}}{2\pi}~
    t_{\mathrm{ggg}\kappa}(\ell_1,\ell_2,\ell_3,\psi_{12},\psi_{23})~
    f_\mathrm{ggg\kappa}(\ell_1,\ell_2,\ell_3,\psi_{12},\psi_{23};\vartheta_1,\vartheta_2,\vartheta_3,\phi_{12},\phi_{23}) \;, \\
    &t_{\mathrm{ggg}\kappa}(\ell_1,\ell_2,\ell_3,\psi_{12},\psi_{23}) 
    &&= 2\pi \left(\prod\limits_{i=1}^3 \int_0^\infty \dd\vartheta_i~\vartheta_i\right)\int_0^{2\pi}\dd\phi_{12}\int_0^{2\pi}\dd\phi_{23}~
    \nnngCF(\vartheta_1,\vartheta_2,\vartheta_3,\phi_{12},\phi_{23})~f^*_\mathrm{ggg\kappa}(\ell_1,\ell_2,\ell_3,\psi_{12},\psi_{23};\vartheta_1,\vartheta_2,\vartheta_3,\phi_{12},\phi_{23})\;,
\end{alignedat}
\ee
where `$^*$' denotes the complex conjugate and the filter function is defined as 
\be
\begin{alignedat}{1}
f_\mathrm{ggg\kappa}(\ell_1,\ell_2,\ell_3,\psi_{12},\psi_{23};\vartheta_1,\vartheta_2,\vartheta_3,\phi_{12},\phi_{23}) 
&= \left[\eexp^{2\iunit\nu}~J_2(A)\right](\ell_1\vartheta_1,\ell_2\vartheta_2,\ell_3\vartheta_3,\phi_{12}-\psi_{12},\phi_{23}-\psi_{23})~
\eexp^{2\iunit\nu(\ell_1,\ell_2,\ell_3,\psi_{12},\psi_{23})}.
\end{alignedat}
\ee

\subsection{\label{app:A_apstats} Derivation of the fourth-order aperture statistics}
In the following, we will derive an explicit expression for the filter function $A_{\mathcal{N}^3M}$ appearing in Eq. \eqref{eq:N3M_final}. 
Based on the calculations presented here, the filter functions for the remaining fourth-order aperture statistics can be derived in a similar fashion. 
The derivation below is done in a similar fashion to the calculation of the filter funtion for the $\langle\Nap M^2\rangle$ statistics presented in the appendix of \cite{Schneider_2005}.

In order to avoid confusion of the aperture scales $\theta_1, \theta_2, \theta_3$ and $\theta_4$ with the galaxy position vectors $\pmb{\theta}_1, \pmb{\theta}_2, \pmb{\theta}_3$ and $\pmb{\theta}_4$, as they were used in Fig. \ref{fig:gggs_sketch}, we change the latter to $\pmb{X}_1, \pmb{X}_2, \pmb{X}_3$ and $\pmb{Y}$ in the following calculations.
To arrive at Eq. \eqref{eq:N3M_final}, we need to start from the basic definition of the aperture statistics that reads 
\be
\label{eq:app_N3M_1}
\begin{alignedat}{1}
    \langle\mathcal{N}^3 M\rangle(\theta_1,\theta_2,\theta_3;\theta_4) 
    &= \left(\prod\limits_{i=1}^3 \int_{\mathbb{R}^2} \dd^2X_i~U_{\theta_i}(|\pmb{X}_i|)\right) \int_{\mathbb{R}^2}\dd^2Y~Q_{\theta_4}(|\pmb{Y}|) 
    \langle\kappa_\mathrm{g}(\pmb{X}_1)~\kappa_\mathrm{g}(\pmb{X}_2)~\kappa_\mathrm{g}(\pmb{X}_3)~\gamma(\pmb{Y};\psi)\rangle \\
    &= \frac{1}{(4\pi\theta_1\theta_2\theta_3\theta_4)^4} \left(\prod\limits_{i=1}^3 \int_{\mathbb{R}^2} \dd^2\vartheta_i\right) \eexp^{2\iunit\zeta}~ \nnngCF(\pmb{\vartheta}_1,\pmb{\vartheta}_2,\pmb{\vartheta}_3) \\
    &\hspace{2.5cm} \times \int_{\mathbb{R}^2}\dd^2Y~(\breve{Y}^*)^2 \left(\prod\limits_{i=1}^3 \left(2\theta_i^2-|\pmb{Y}+\pmb{\vartheta}_i|^2\right)\right)
    \exp\left[-\frac{1}{2}\left( \frac{|\pmb{Y}|^2}{\theta_4^2} + \sum_{i=1}^{3} \frac{|\pmb{Y}+\pmb{\vartheta}_i|^2}{\theta_i^2} \right)\right]\;,
\end{alignedat}
\ee
where we have already inserted the expressions of the $U_\theta$ and $Q_\theta$ filter functions (see Eqs. \ref{eq:filter_functions_definition}). 
In the first line, we denote the polar angle of the vector $\pmb{Y}$ with $\psi$. 
Applying the complex notation $\breve{Y}=Y_1+\iunit Y_2=|\pmb{Y}|~\eexp^{\iunit\psi}$ for the vector $\pmb{Y}$, one can write the rotated shear as 
\be
\gamma(\pmb{Y};\psi) = \gamma(\pmb{Y};\zeta)~\eexp^{2\iunit(\zeta - \psi)} 
= \gamma(\pmb{Y};\zeta)~\eexp^{2\iunit\zeta} \frac{(\breve{Y}^*)^2}{|\pmb{Y}|^2}\;.
\ee
Substituting the separation vectors $\pmb{\vartheta}_i=\pmb{X}_i-\pmb{Y}$ with polar angles $\varphi_i$, we can use the definition of $\nnngCF$ in Eq. \eqref{eq:Ggggk_definition} to arrive at the second line in Eq. \eqref{eq:app_N3M_1}.
For the final expression, we should obtain an integral over the parameter space $(\vartheta_1,\vartheta_2,\vartheta_3,\phi_{12},\phi_{23})$ of the 4PCF, which means that we need to carry out three further integrations in Eq. \eqref{eq:app_N3M_1}. 
The expression in the exponential function can be rewritten as
\begin{equation}
    \sum_{i=1}^{3} \frac{|\pmb{Y}+\pmb{\vartheta}_i|^2}{2\theta_i^2} + \frac{|\pmb{Y}|^2}{2\theta_4^2}
    = \frac{|\pmb{y}|^2}{a} + b\;,
    \quad\text{with}\quad a = \frac{2}{\sum_{i=1}^4 \theta_i^{-2}} ~,\quad 
    b = \frac{1}{2}\sum\limits_{i=1}^3\frac{|\pmb{\vartheta}_i|^2}{\theta_i^2} - \frac{|\pmb{c}|^2}{a}\;,
\end{equation}
where we apply the substitution $\pmb{Y}=\pmb{y}-\pmb{c}$, i.e. a translation of $\pmb{Y}$ by the vector 
\begin{equation}
    \pmb{c} = \frac{a}{2}\sum\limits_{i=1}^3 \frac{\pmb{\vartheta}_i}{\theta_i^2}\;.
\end{equation}
We further define the complex quantity $\breve{g}_i = |\pmb{\vartheta}_i| - \breve{c}^*\eexp^{\iunit\varphi_i}$, such that in the prefactor of the $\pmb{Y}$-integration in Eq. \eqref{eq:app_N3M_1}, we can write
\be
    |\pmb{Y}+\pmb{\vartheta}_i|^2 
    = |\breve{y}+\breve{g}_i^*~\eexp^{\iunit\varphi_i}|^2
    = |\breve{y}|^2 + |\breve{g}_i|^2 + \breve{y}\breve{g}_i^*~\eexp^{\iunit\varphi_i} + \breve{y}^*\breve{g}_i~\eexp^{-\iunit\varphi_i} \;.
\ee
With this, we can express the whole prefactor by 
\be
\begin{alignedat}{1}
    F_{\Nap^3M} &= (\breve{Y}^*)^2~\eexp^{2\iunit\zeta}\left(\prod\limits_{i=1}^3 \left(2\theta_i^2-|\pmb{Y}+\pmb{\vartheta}_i|^2\right)\right) \\
    &= (\breve{y}^*-\breve{c}^*)^2~\eexp^{2\iunit\zeta}
    \left(\prod\limits_{i=1}^3 \left(2\theta_i^2 - |\breve{y}|^2 - |\breve{g}_i|^2 - \breve{y}\breve{g}_i^*~\eexp^{\iunit\varphi_i} - 
    \breve{y}^*\breve{g}_i~\eexp^{-\iunit\varphi_i}\right)\right) \\
    &= (\breve{c}^*~\eexp^{\iunit\zeta})^2 \bigg[2\Re\left[\breve{g}_1\breve{g}_2^* \eexp^{\iunit\phi_{12}}\right](F_3|\pmb{y}|^2-|\pmb{y}|^4) +
    2\Re\left[\breve{g}_1\breve{g}_3^* \eexp^{\iunit(\phi_{12}+\phi_{23})}\right](F_2|\pmb{y}|^2-|\pmb{y}|^4)
    + 2\Re\left[\breve{g}_2\breve{g}_3^* \eexp^{\iunit\phi_{23}}\right](F_1|\pmb{y}|^2-|\pmb{y}|^4) \\
    &\qquad\qquad + F_1 F_2 F_3 - |\pmb{y}|^2(F_1F_2 + F_1F_3 + F_2F_3) + |\pmb{y}|^4(F_1+F_2+F_3) - |\pmb{y}|^6\bigg] \\
    &\quad +2\breve{c}^*~\eexp^{\iunit\zeta}\bigg[\breve{g}_1\eexp^{\iunit(\zeta-\varphi_1)} \left(|\pmb{y}|^4\Re\left[\breve{g}_2\breve{g}_3^* \eexp^{\iunit\phi_{23}}\right] + F_2F_3|\pmb{y}|^2 - |\pmb{y}|^4(F_2+F_3) + |\pmb{y}|^6 \right) \\
    &\qquad\qquad + \breve{g}_2\eexp^{\iunit(\zeta-\varphi_2)} \left(|\pmb{y}|^4\Re\left[\breve{g}_1\breve{g}_3^* \eexp^{\iunit(\phi_{12}+\phi_{23})}\right] + |\pmb{y}|^2F_1F_3 - |\pmb{y}|^4(F_1+F_3) + |\pmb{y}|^6 \right) \\
    &\qquad\qquad + \breve{g}_3\eexp^{-\iunit(\zeta-\varphi_3)} \left(|\pmb{y}|^4\Re\left[\breve{g}_1\breve{g}_2^* \eexp^{\iunit\phi_{12}}\right] + |\pmb{y}|^2F_1F_2 - |\pmb{y}|^4(F_1+F_2) + |\pmb{y}|^6\right) \bigg] \\
    &\quad + \breve{g}_1\breve{g}_2\eexp^{\iunit(2\zeta-\varphi_1-\varphi_2)}(F_3|\pmb{y}|^4-|\pmb{y}|^6)
    + \breve{g}_1\breve{g}_3\eexp^{\iunit(2\zeta-\varphi_1-\varphi_3)}(F_2|\pmb{y}|^4-|\pmb{y}|^6) + \breve{g}_2\breve{g}_3\eexp^{\iunit(2\zeta-\varphi_2-\varphi_3)}(F_1|\pmb{y}|^4-|\pmb{y}|^6)
    + C\;,
\end{alignedat}
\ee
where we defined $F_i=2\theta_i^2-|\breve{g}_i|^2$ and collected all the terms that vanish upon the integration over the angular part of $\pmb{y}$, in $C$. 
In the $\pmb{y}$-integration, we encounter integrals of the type
$\int\dd y~y^n~\eexp^{-y^2/a} = \frac{1}{2}(\frac{n-1}{2})!~a^{(n+1)/2}$, for odd integers $n$.

\begin{figure}[t]
    \centering
    \includegraphics[width=\textwidth]{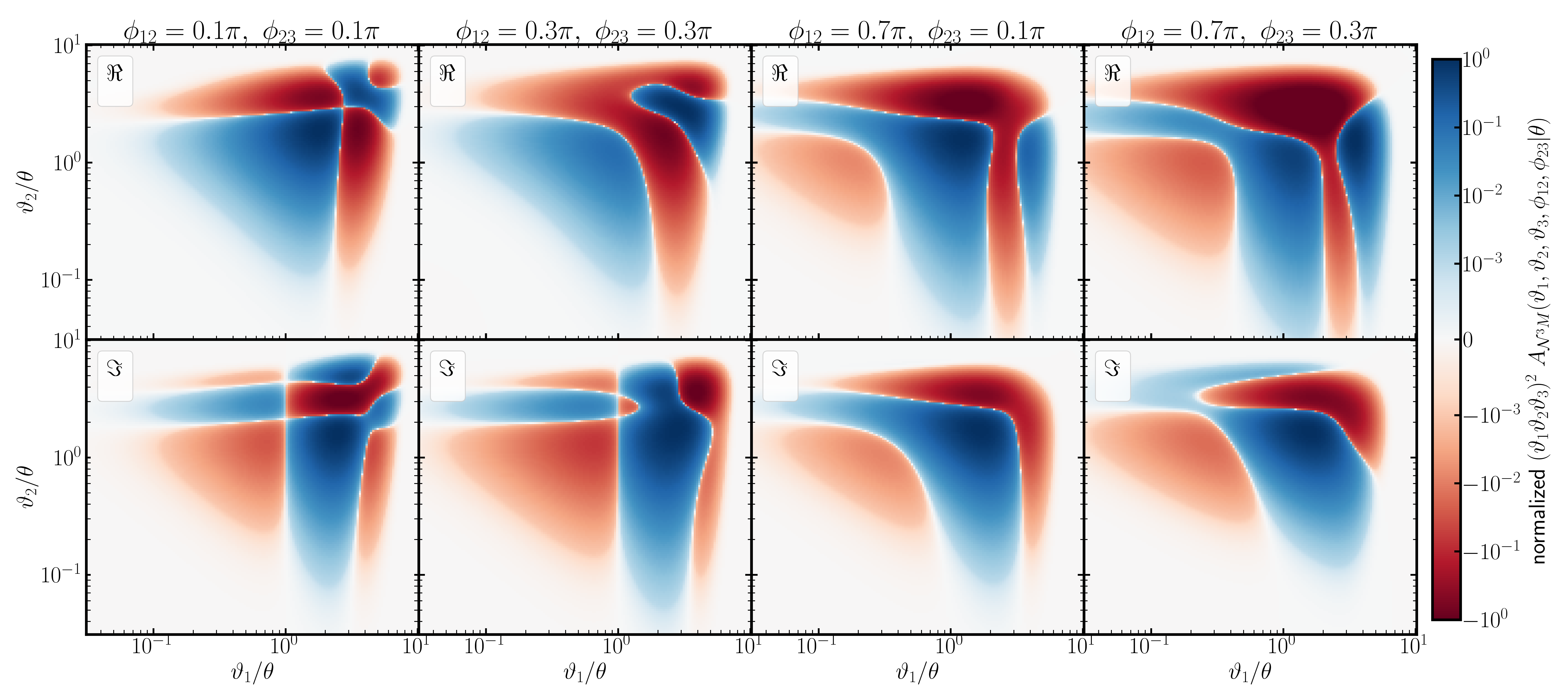}
    \caption{The filter function $(\vartheta_1 \vartheta_2 \vartheta_3)^2 A_{\mathcal{N}^3M}(\vartheta_1,\vartheta_2,\vartheta_3,\phi_{12},\phi_{23}|\theta)$ as a function of $\vartheta_{1,2}/\theta$ on a logarithmic scale for four different configurations of $(\phi_{12},\phi_{23})$. 
    The top plots show the real part and the bottom plots the imaginary part of the function.
    All aperture scales are fixed to $\theta=5'$, as well as the third radial coordinate $\vartheta_3=\theta$. 
    The values of the color scale are spaced logarithmically for both positive and negative values with a linear domain for values in the interval $[-10^{-3},10^{-3}]$, and all values are normalized by the maximum value.}
    \label{fig:AN3M_pcolor}
\end{figure}

In the angular part of the $\pmb{\vartheta}_i$-integrals in Eq. \eqref{eq:app_N3M_1}, we apply the transformation 
$\zeta=\frac{\varphi_1+\varphi_3}{2}$, $\phi_{12}=\varphi_2-\varphi_1$, $\phi_{23}=\varphi_3-\varphi_2$, such that the dependencies of 
the 4PCF are consistent with the previous definitions. The Jacobi determinant of this transformation is $1$ 
and the integration over $\zeta$ is trivial, yielding a factor of $2\pi$. 
We then obtain the expression of the filter function to 
\be
\label{eq:app_AN3M}
\begin{alignedat}{1}
    A_{\mathcal{N}^3M}(\vartheta_1,\vartheta_2,\vartheta_3,\phi_{12},&\phi_{23}|\theta_1,\theta_2,\theta_3;\theta_4) \\
    = \frac{\eexp^{-b}}{512\pi^2\Theta^{12}} \Bigg\{&\breve{h}^2
    \bigg[2\Re\left[\breve{g}_1\breve{g}_2^* \eexp^{\iunit\phi_{12}}\right](F_3-2a) + 2\Re\left[\breve{g}_1\breve{g}_3^* \eexp^{\iunit(\phi_{12}+\phi_{23})}\right](F_2-2a) 
    + 2\Re\left[\breve{g}_2\breve{g}_3^* \eexp^{\iunit\phi_{23}}\right](F_1-2a) \\
    &\hspace{0.4cm} + \frac{F_1 F_2 F_3}{a} - (F_1F_2 + F_1F_3 + F_2F_3) + 2a(F_1+F_2+F_3-3a)\bigg] \\
    &+ 2\breve{h} \bigg[\breve{g}_1\eexp^{\iunit(\phi_{12}+\phi_{23})/2} \left(2a\Re\left[\breve{g}_2\breve{g}_3^* \eexp^{\iunit\phi_{23}}\right] + F_2F_3 - 2a(F_2+F_3-3a) \right) \\
    &\hspace{0.8cm} + \breve{g}_2\eexp^{\iunit(\phi_{23}-\phi_{12})/2} \left(2a\Re\left[\breve{g}_1\breve{g}_3^* \eexp^{\iunit(\phi_{12}+\phi_{23})}\right] + F_1F_3 - 2a(F_1+F_3-3a) \right) \\
    &\hspace{0.8cm} + \breve{g}_3\eexp^{-\iunit(\phi_{12}+\phi_{23})/2} \left(2a\Re\left[\breve{g}_1\breve{g}_2^* \eexp^{\iunit\phi_{12}}\right] + F_1F_2 - 2a(F_1+F_2-3a) \right)\bigg] \\
    &+ 2a \bigg[\breve{g}_1\breve{g}_2\eexp^{\iunit\phi_{23}}(F_3-3a)
    + \breve{g}_1\breve{g}_3(F_2-3a) + \breve{g}_2\breve{g}_3\eexp^{-\iunit\phi_{12}}(F_1-3a) \bigg]\Bigg\}\;,
\end{alignedat}
\ee
where we further defined 
\be
    \breve{h} = \breve{c}^*\eexp^{\iunit\zeta} = \frac{a}{2}\left(\frac{\vartheta_1 \eexp^{\iunit(\phi_{12}+\phi_{23})/2}}{\theta_1^2}
    + \frac{\vartheta_2 \eexp^{\iunit(\phi_{23}-\phi_{12})/2}}{\theta_2^2}
    + \frac{\vartheta_3 \eexp^{-\iunit(\phi_{12}+\phi_{23})/2}}{\theta_3^2}\right)\,,
\quad\text{and}\quad
\Theta^6 = \frac{1}{4}\left(\theta_1^2\theta_2^2\theta_3^2 + \theta_1^2\theta_2^2\theta_4^2 + \theta_1^2\theta_3^2\theta_4^2 + \theta_2^2\theta_3^2\theta_4^2\right)\,.
\ee
The quantity $b$ can be written in terms of all integration variables as 
\be
b = \frac{1}{2}\sum\limits_{i=1}^3\frac{\vartheta_i^2}{\theta_i^2} \left(1-\frac{a}{2\theta_i^2}\right)
    - \frac{a}{4}\left[\frac{2\vartheta_1\vartheta_2 \cos\phi_{12}}{\theta_1^2\theta_2^2}
    + \frac{2\vartheta_1\vartheta_3 \cos\left(\phi_{12}+\phi_{23}\right)}{\theta_1^2\theta_3^2}
    + \frac{2\vartheta_2\vartheta_3 \cos\phi_{23}}{\theta_2^2\theta_3^2}\right]\,.
\ee
In Fig. \ref{fig:AN3M_pcolor} this filter function is plotted for a number of different configurations.

%

\section{\label{app:numint} Second-order aperture statistics}
If we assume Gaussian shear and galaxy fields, the 4PCF, as defined in Sect. \ref{ssc:4PCF}, can be decomposed into products of 2PCFs according to 
\be
\label{eq:Ggggk_gaussian_fields}
\begin{alignedat}{1}
    G_{\mathrm{ggg}\kappa}(\vartheta_1,\vartheta_2,\vartheta_3,\phi_{12},\phi_{23}) = 
    \langle\kappa_\mathrm{g} \kappa_\mathrm{g}\rangle(\Delta\theta_{12}) \langle\kappa_\mathrm{g} \gamma\rangle(\vartheta_3)~\eexp^{\iunit(\phi_{12}+\phi_{23})} 
    + \langle\kappa_\mathrm{g} \kappa_\mathrm{g}\rangle(\Delta\theta_{13}) \langle\kappa_\mathrm{g} \gamma\rangle(\vartheta_2)~\eexp^{\iunit(\phi_{12}-\phi_{23})} 
    + \langle\kappa_\mathrm{g} \kappa_\mathrm{g}\rangle(\Delta\theta_{23}) \langle\kappa_\mathrm{g} \gamma\rangle(\vartheta_1)~\eexp^{-\iunit(\phi_{12}+\phi_{23})}\;,
\end{alignedat}
\ee
with $\Delta\theta_{ij} = |\pmb{\vartheta}_i-\pmb{\vartheta}_j| = \sqrt{\vartheta_i^2 + \vartheta_j^2 - 2\vartheta_i\vartheta_j \cos\phi_{ij}}$. 

In order to compute the expected results of the fourth-order aperture statistics $\langle\Nap^3\Map\rangle$ in the case of Gaussian fields, we need the second-order aperture statistics $\langle\Nap^2\rangle$ and $\langle\Nap\Map\rangle$. 
Both quantities can be expressed by one-dimensional integrals over the respective 2PCF.

In the case of $\langle\Nap\Map\rangle$, the corresponding 2PCF is given in Eq. \eqref{eq:2ndorder_gscorr} and is only a function of the separation $\vartheta$. The integral reads 
\be
\langle\Nap M\rangle(\theta) 
= - \int_{\mathbb{R}^2}\dd^2 X~U_\theta(|\pmb{X}|) \int_{\mathbb{R}^2}\dd^2 Y~Q_\theta(|\pmb{Y}|) \langle\kappa_\mathrm{g}(\pmb{X}) \gamma(\pmb{Y})\rangle 
\frac{(\breve{Y}^*)^2}{|\pmb{Y}|^2}
= \int_0^\infty\frac{\dd\vartheta~\vartheta}{\theta^2} T_\times\left(\frac{\vartheta}{\theta}\right) \langle\gamma_\mathrm{t}\rangle(\vartheta)\;,
\ee
where the filter function is defined as
\be
T_\times(s) = s^2\frac{12-s^2}{128}\eexp^{-s^2/4}\;,
\ee
in accordance with the aperture mass filter functions $T_\pm$ in \cite{Jarvis_2004}. 
The $\langle\Nap\Map\rangle$ statistics with two different aperture scales does not contain new information, as this can be computed via 
$\langle\Nap\Map\rangle(\theta_1,\theta_2)=\frac{\theta_1^2\theta_2^2}{\theta^4}\langle\Nap\Map\rangle(\theta)\Bigg|_{\theta=\sqrt{(\theta_1^2+\theta_2^2)/2}}$.

The $\langle\Nap^2\rangle$ statistics can be computed via an integral over the 2PCF $\langle\kappa_\mathrm{g}\kappa_\mathrm{g}\rangle$ that reads
\be
\langle\Nap^2\rangle(\theta) 
= \int_{\mathbb{R}^2}\dd^2 X~U_\theta(|\pmb{X}|) \int_{\mathbb{R}^2}\dd^2 Y~U_\theta(|\pmb{Y}|) \langle\kappa_\mathrm{g}(\pmb{X}) \kappa_\mathrm{g}(\pmb{Y})\rangle
= \int_0^\infty\frac{\dd\vartheta~\vartheta}{\theta^2} T_+\left(\frac{\vartheta}{\theta}\right) \langle\kappa_\mathrm{g}\kappa_\mathrm{g}\rangle(\vartheta)\;,
\ee
where the filter function is given as \citep{Jarvis_2004}
\be
T_+(s) = \frac{s^4-16s^2+32}{128}\eexp^{-s^2/4}\;.
\ee
Again, for different aperture scales, we have 
$\langle\Nap^2\rangle(\theta_1,\theta_2) = \frac{\theta_1^2\theta_2^2}{\theta^4}\langle\Nap^2\rangle(\theta)\Bigg|_{\theta=\sqrt{(\theta_1^2+\theta_2^2)/2}}$.

The expected result for the fourth-order aperture statistics can then be obtained via 
\be
\langle\Nap^3\Map\rangle(\theta_1,\theta_2,\theta_3;\theta_4) 
= \langle\Nap^2\rangle(\theta_1,\theta_2)\langle\Nap\Map\rangle(\theta_3,\theta_4) 
+\langle\Nap^2\rangle(\theta_1,\theta_3)\langle\Nap\Map\rangle(\theta_2,\theta_4)
+\langle\Nap^2\rangle(\theta_2,\theta_3)\langle\Nap\Map\rangle(\theta_1,\theta_4)\;.
\ee

\section{\label{app:multipole_est}Multipole based estimator for the 4PCF}
Given a survey consisting of $N_{\rm{sources}}$ source galaxies and $N_{\rm{lens}}$ lenses, and a binning scheme of radial and angular bins, the traditional estimator of the bin-averaged G4L correlation function assigns all source-lens-lens-lens quadruplets to their corresponding quadrilateral configuration bin and then averages over those:
\begin{align}\label{eq:Gtilde_quadrupletestimator}
\tilde{G}&_{\mathrm{ggg}\gamma}(\vartheta_1,\vartheta_2,\vartheta_3,\phi_{12},\phi_{13})
\equiv \frac{\tilde{G}'_{\mathrm{ggg}\gamma}(\vartheta_1,\vartheta_2,\vartheta_3,\phi_{12},\phi_{13})}{N(\vartheta_1,\vartheta_2,\vartheta_3,\phi_{12},\phi_{13})} \ ,
\end{align}
where the $\vartheta_i$ and the $\phi_{ij}$ denote the radial and angular bins.
To achieve a worst-case quadratic time complexity of the estimator we use its representation in the multipole basis, where the angular arguments of the 4PCF are expanded in complex exponentials. Adopting the notation of \citet{Porth_2025}, we denote the 4PCF multipoles as $\tilde{G}_{\mathrm{ggg}\gamma,\mathbf{n}}$, where $\mathbf{n} \equiv \left(n_2, n_3\right)$. In general, the two bases are related as
\begin{align}\label{eq:Multipole2RealConversion}
\tilde{G}'_{\mathrm{ggg}\gamma}\left(\vartheta_1,\vartheta_2,\vartheta_3,\phi_{12},\phi_{13}\right) \equiv
     \frac{1}{(2\pi)^2} \sum_{n_2,n_3=-\infty}^\infty \tilde{G}'_{\mathrm{ggg}\gamma,\mathbf{n}}(\vartheta_1,\vartheta_2,\vartheta_3) \ \mathrm{e}^{\mathrm{i} n_2\phi_{12}}\mathrm{e}^{\mathrm{i} n_{3}\phi_{13}} \ ,
\end{align}
and the multipole components can be obtained as
\begin{align}
    \tilde{G}'_{\mathrm{ggg}\gamma,\mathbf{n}}\left(\vartheta_1,\vartheta_2,\vartheta_3\right)  &= \sum_i^{N_{\mathrm{source}}} w_{\mathrm{source}}(\pmb{\theta}_i) \, \gamma_{\mathrm c}(\pmb{\theta}_i) \ W^{\mathrm{disc}}_{n_2+n_3-1}(\pmb{\theta}_i,\vartheta_1) 
    \, W^{\mathrm{disc}}_{-n_2}(\pmb{\theta}_i,\vartheta_2) \ \, W^{\mathrm{disc}}_{-n_3-1}(\pmb{\theta}_i,\vartheta_3)
    \ , \\ 
     W^{\mathrm{disc}}_{n}(\pmb{\theta_i},\vartheta) &\equiv \sum_{k=1}^{N_{\mathrm{lens}}} w_{\mathrm{lens}}(\pmb{\theta}_k) \, \mathrm{e}^{\mathrm{i} n \varphi_{ik}} \ \mathcal{B}(\theta_{ik} \in \vartheta) \ ,
\end{align}
where we defined $\pmb{\theta}_k-\pmb{\theta}_i \equiv \theta_{ik} \mathrm{e}^{\mathrm{i}\varphi_{ik}}$ and we further introduced the bin selection function $\mathcal{B}(x\in X)$, 
which is unity if $x\in X$ and zero otherwise. For the corresponding expressions of the normalization we refer to \citet{Porth_2025}. 

We implement the multipole-based G4L estimator in the \textsc{orpheus} software package\footnote{\url{https://github.com/lporth93/orpheus}}, where we achieve a further acceleration by using tree-based methods for the allocation of the $W_n$. We also take into account the multiple-counting corrections and make use of the symmetry properties of the $\tilde{G}'_{\mathrm{ggg}\gamma,\mathbf{n}}$, both of which can be obtained using the same prescription as given in appendix A of \citet{Porth_2025}.


\section{Estimator for aperture statistics}
\subsection{\label{app:directestimator} Direct estimator}
Starting from the estimator in Eq. \eqref{eq:Mapm_Napn_estimator} for the mixed statistics $\widehat{\Nap^n \Map^m}$ with arbitrary order $n$ in $\Nap$ and $m$ in $\Map$, we first note that this expression factorizes into the $\Nap^n$--part and the $\Map^m$--part, because we sum over different galaxy populations, in general. 
In the following, we focus on equal aperture scales in each $\Nap$ and $\Map$ but distinguish between $\thap{N}{}$ and $\thap{M}{}$. 

We can decompose the sums over all galaxies inside an aperture of area $\Aap(\theta_\mathrm{N})$, as given in Eq. \eqref{eq:Mapm_Napn_estimator}, into sums that are linear in the number of galaxies. In the case of $n=2$, we can decompose the sum according to 
\be
    \widehat{\Nap^2}(\theta_\mathrm{N}) = \Aap^2(\theta_\mathrm{N}) \frac{\sum_{i_1\neq i_2} U_{\thap{N}{};i_1} U_{\thap{N}{};i_2}}{\sum_{i_1\neq i_2} 1} 
    = \Aap^2(\theta_\mathrm{N}) \frac{\sum_{i_1,i_2} U_{\thap{N}{};i_1}U_{\thap{N}{};i_2} - \sum_i U_{\thap{N}{};i}^2}{n_\mathrm{L}(n_\mathrm{L}-1)} 
    = \frac{\Naps{1}^2 - \Naps{2}}{1-1/n_\mathrm{L}}\;,
\ee
where we defined the quantity 
\be
\label{eq:Napsm}
    \Naps{k}(\thap{N}{}) = \frac{1}{\Nlens^k} \sum_i U_{\thap{N}{};i}^k\;, 
\ee
with the lens galaxy number density $\Nlens = n_\mathrm{L}/\Aap(\theta_\mathrm{N})$, and $n_\mathrm{L}$ the number of lens galaxies inside the aperture. 
All further higher-order $\widehat{\Nap^n}$ statistics can be expressed in terms of the $\Naps{k}$ with $k=1,\dots,n$. 
The general $m$-th order aperture mass can be computed recursively using the complete Bell polynomials $B_n$ as was shown in \cite{Porth_2021}. Similarly, the $n$-th order moment $\Nap^n$ is estimated by 
\be
\label{eq:Napn_Bell}
\widehat{\Nap^n} = \frac{B_n\left(-\Naps{1},\dots,-(n-1)!\Naps{n}\right)}{\prod_{k=0}^{n-1} (1-k/n_\mathrm{L})}\;.
\ee
A crude approximation of $\widehat{\Nap^n}$ is to set $\widehat{\Nap^n}\approx\Naps{1}^n(\thap{N}{})$ which neglects the subsequent correction terms including $\Naps{k}(\thap{N}{})$ with $k>1$.
While for large aperture scales this approximation can yield sufficiently precise results, it is only a poor estimate on small scales where the impact of the correction terms becomes more significant. 
Using the global number density instead of the local one inside the respective aperture as the normalization in Eq. \eqref{eq:Napsm}, changes results for $\widehat{\Nap^2}$ already by a few per-cent, because we fail to account for fluctuations in the galaxy counts -- an effect most prominent for small apertures. 
This is similar to what was described for the third-order aperture mass $\langle\Map^3\rangle$ in \cite{Heydenreich_2023}. 
In general, we find a better fit with other methods, as shown in Sect. \ref{ssc:comparison_estimators}, using the local number density. 

For the $\widehat{\Map^m}$ statistics, we adopt the notation from \cite{Porth_2021} and write 
\be
\label{eq:Mapsm}
\Maps{k}(\thap{M}{}) = \Aap^k(\thap{M}{}) S_k \frac{\sum_i \left(w^\mathrm{S}_i Q_{\thap{M}{};i} \epsilon_{\mathrm{t},i} \right)^k}{\sum_i \left(w^\mathrm{S}_i \right)^k}, \quad \text{with}
\quad S_k = \frac{\sum_i \left(w^\mathrm{S}_i \right)^k}{\left(\sum_i w^\mathrm{S}_i \right)^k}\;,
\ee
such that we can calculate the aperture mass moments via 
\be
\label{eq:Mapm_Bell}
\widehat{\Map^m} = \frac{B_m\left(-\Maps{1},\dots,-(m-1)!\Maps{m}\right)}{B_m\left(-S_1,\dots,-(m-1)!S_m\right)}\;.
\ee

\subsection{\label{app:directestimator_discrete} Discretization of the direct estimator}
In order to estimate aperture moments from discretized data, we need to convert the sums over the individual galaxies in the definitions of $\Naps{k}$ and $\Maps{k}$ into sums over all pixels inside an aperture. 
For the following calculations, we split up the sum over the galaxies according to 
\be
    \sum_\mathrm{gal}1 = \sum_\mathrm{pix} \sum_\mathrm{gal\in pix}1 = \sum_\mathrm{pix} n_\mathrm{pix}\,,
\ee
where the sum $n_\mathrm{pix}=\sum_\mathrm{gal\in pix}$ denotes the number of galaxies, $n_\mathrm{pix}$, in the corresponding pixel obtained from the sum  over all galaxies within the pixel. 
With these ingredients, we can transform the definition for $\Naps{k}$ to the pixel level by writing 
\be
\label{eq:Nsm_pixel_estimator}
    \Naps{k}(\thap{N}{}) = \frac{1}{\Nlens^k} \sum_\mathrm{gal} U_{\thap{N}{};\mathrm{gal}}^k
    \approx \frac{1}{\Nlens^k} \sum_\mathrm{pix} U_{\thap{N}{};\mathrm{pix}}^k \sum_\mathrm{gal \in pix} 1 
    = \frac{1}{\Nlens^k} \sum_\mathrm{pix} U_{\thap{N}{};\mathrm{pix}}^k n_\mathrm{L,pix}\;,
\ee
where we denote the number of lens galaxies in a pixel by $n_\mathrm{L,pix}$. 
The $\Naps{k}$ can thus be simply computed by summing up the product of $n_\mathrm{L,pix}$ with the corresponding $k$-th moment of the filter function $U_{\thap{N}{};\mathrm{pix}}$, evaluated at the pixel center, over all pixels in an aperture. 

In order to convert the $\Maps{k}$ to the pixel level, we first define the $k$-th order ellipticity weight for a pixel as 
\be
    w^\mathrm{S}_{k,\mathrm{pix}} = \sum_\mathrm{gal \in pix} \left(w^\mathrm{S}_\mathrm{gal} \right)^k, 
\ee
which then allows us to write 
\be
\label{eq:Msm_pixel_estimator}
\begin{alignedat}{1}
    \Maps{k}(\thap{M}{}) &= \Aap^k(\theta_\mathrm{M}) S_k \frac{\sum_\mathrm{gal} \left(w^\mathrm{S}_\mathrm{gal} Q_{\thap{M}{};\mathrm{gal}} \epst{,\mathrm{gal}} \right)^k}{\sum_\mathrm{gal} \left(w^\mathrm{S}_\mathrm{gal} \right)^k} 
    \approx \Aap^k(\theta_\mathrm{M}) S_k \frac{\sum_\mathrm{pix} Q_{\thap{M}{};\mathrm{pix}}^k \sum_\mathrm{gal\in pix} \left(w^\mathrm{S}_\mathrm{gal} \epst{,\mathrm{gal}} \right)^k}{\sum_\mathrm{pix} w^\mathrm{S}_{k,\mathrm{pix}}} 
    = \Aap^k(\theta_\mathrm{M}) S_k \frac{\sum_\mathrm{pix} w^\mathrm{S}_{k,\mathrm{pix}} Q_{\thap{M}{};\mathrm{pix}}^k \epsilon_{\mathrm{t,pix},k}}{\sum_\mathrm{pix} w^\mathrm{S}_{k,\mathrm{pix}}}\;,
\end{alignedat}
\ee
where we further defined $\epsilon_{\mathrm{t,pix},k}$ as the mean $k$-th ellipticity moment in a pixel given by 
\be
    \epsilon_{\mathrm{t,pix},k} = \frac{\sum_\mathrm{gal \in pix} \left(w^\mathrm{S}_\mathrm{gal}\epsilon_{\mathrm{t,gal}} \right)^k}{w^\mathrm{S}_{k,\mathrm{pix}}}\;.
\ee
The summation over all pixels inside an aperture can be performed efficiently by convolving the discretized shear or galaxy number density field with the corresponding filter function, which is also discretized on the same pixel grid as the data.

\end{appendix}

\label{LastPage}
\end{document}